\pdfoutput=1
\documentclass[11pt,twoside,a4paper,cmspaper,final,collab]{cms-tdr}
\def\svnVersion{5abee16}\def\svnDate{2026/08/03}\def\cmsCernNoTag{CERN-EP-2025-299}\def\cmsCernDate{\today}\def\cmsMessage{Published in the European Physical Journal C as \href{http://dx.doi.org/10.1140/epjc/s10052-026-15699-9}{\doi{10.1140/epjc/s10052-026-15699-9}.}}
\begin{document}\cmsNoteHeader{B2G-24-014}

\hyphenation{had-ron-i-za-tion}
\hyphenation{cal-or-i-me-ter}
\hyphenation{de-vices}
\ifthenelse{\boolean{cms@external}}{\providecommand{\cmsLeft}{upper\xspace}}{\providecommand{\cmsLeft}{left\xspace}}
\ifthenelse{\boolean{cms@external}}{\providecommand{\cmsRight}{lower\xspace}}{\providecommand{\cmsRight}{right\xspace}}

\newcommand{\pp}{\ensuremath{\Pp\Pp}\xspace}
\newcommand{\tautau}{\ensuremath{\PGtp\PGtm}\xspace}
\newcommand{\HH}{\ensuremath{\PH\PH}\xspace}
\newcommand{\ttjets}{\ensuremath{\ttbar}+jets\xspace}
\newcommand{\MHbb}{\ensuremath{M_{\PH(\bbbar)}}\xspace}
\newcommand{\wjets}{\ensuremath{\PW}+jets\xspace}
\newcommand{\zjets}{\ensuremath{\PZ}+jets\xspace}
\newcommand{\FH}{\ensuremath{\tauh\tauh}\xspace}
\newcommand{\SL}{\ensuremath{\Pell\tauh}\xspace}
\newcommand{\Pelldag}{{\HepParticle{\Pell}{}{\ddag}}\xspace}
\newcommand{\msjone}{\ensuremath{m_\mathrm{subjet\,1}}\xspace}
\newcommand{\msjtwo}{\ensuremath{m_\mathrm{subjet\,2}}\xspace}
\newcommand{\mcajet}{\ensuremath{m_\mathrm{CA8~jet}}\xspace}

\cmsNoteHeader{B2G-24-014}
\titlerunning{Search for heavy resonances decaying into two Higgs bosons}
\title{Search for heavy resonances decaying into two Higgs bosons in the \texorpdfstring{$\bbbar \PGtp \PGtm$}{bbbar tau+ tau-} final state in proton-proton collisions at \texorpdfstring{$\sqrt{s} = 13\TeV$}{sqrt(s) = 13 TeV}}
\date{\today}

\abstract{
A search is presented for massive narrow-width resonances in the mass range of 1--4.5\TeV,  decaying into pairs of Higgs bosons (\HH). The search uses proton-proton collision data at a center-of-mass energy of 13\TeV collected with the CMS detector at the CERN LHC during 2016--2018, corresponding to an integrated luminosity of 138\fbinv. The analysis targets final states where one Higgs boson decays into a pair of bottom quarks and the other into a pair of tau leptons, $\PX \to \HH \to \bbbar \tautau$. It uses a single large radius jet to reconstruct the $\PH\to\bbbar$ decay, while the $\PH\to\tautau$ decay products can either be contained within a single large radius jet or appear as two isolated tau leptons. The observed data are consistent with standard model background expectations. Upper limits at 95\% confidence level are set on the production cross section for resonant \HH production for masses between 1 and 4.5\TeV. This analysis sets the most sensitive limits to date on $\PX \to \HH \to \bbbar \tautau$ decays in the mass range of 1.4 to 4.5\TeV.
}

\hypersetup{
pdfauthor={CMS Collaboration},
pdftitle={Search for heavy resonances decaying into two Higgs bosons in the bbbar tau+ tau- final state in proton-proton collisions at sqrt(s)=13 TeV},
pdfsubject={CMS},
pdfkeywords={CMS, B2G, DIB}
}

\maketitle

\section{Introduction}
\label{sec:introduction}

The discovery of the Higgs boson (\PH) by the CMS~\cite{CMS:2012qbp,CMS:2013btf} and ATLAS~\cite{ATLAS:2012yve,ATLAS:2013xga} experiments confirmed the standard model (SM) as a robust theoretical framework for describing high-energy interactions. However, unresolved questions within the SM, such as the hierarchy problem~\cite{PESKIN2025116971}, have driven the search for beyond-the-SM (BSM) physics.

\begin{figure}[b]
    \centering
    \includegraphics[width=0.5\textwidth]{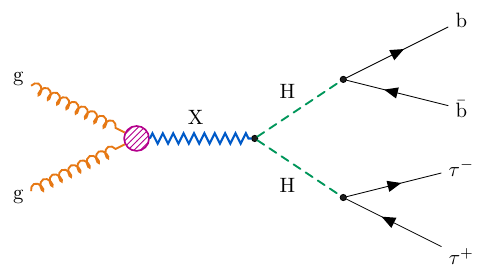}
    \caption{A representative diagram for the production of a spin-0 radion or a spin-2 graviton \PX, which decays into two SM Higgs bosons. One Higgs boson decays into a \bbbar pair and the other into a \tautau pair.}
    \label{fig:intro:simp-model}
\end{figure}

The observation of the \PH boson has opened new avenues for probing BSM scenarios. In fact, many BSM theories predict the existence of heavy resonances (\PX) with masses exceeding 1\TeV that can decay into a pair of Higgs bosons (\HH). The production of Higgs boson pairs is a crucial, but rare, process in the SM. If resonant \HH production occurs through BSM physics, it can significantly enhance the cross section. An example of a well-motivated theoretical framework predicting such enhancements is the warped extra dimensions model proposed by Randall and Sundrum (RS)~\cite{randall1999large, randall1999alternative}, which leads to the existence of new neutral, spin-0 radions ~\cite{goldberger1999modulus, dewolfe2000modeling, csaki2000cosmology} and spin-2 Kaluza--Klein gravitons~\cite{davoudiasl2000phenomenology, agashe2007warped}. The analysis presented in this paper is designed to be model independent, with a single strategy for both spin-0 and spin-2 resonances, each decaying into a pair of Higgs bosons in the $\bbbar\tautau$ final state. A representative diagram illustrating the production of such a resonance is shown in Fig.~\ref{fig:intro:simp-model}. 

{\tolerance=1200 This search targets resonance masses between 1 and 4.5\TeV using the CMS
proton-proton (\pp) collision data sample collected at a center-of-mass energy of $13\TeV$ during the years
2016--2018. A previous CMS search for heavy resonant \HH production in the same
final state~\cite{2016cmsheavybbtautau} is based on a smaller data set, using
only the 2016 \pp collision data, corresponding to 35.9\fbinv.
The observed data in that analysis are consistent with SM expectations, and upper
limits are set at 95\% confidence level (\CL) on the production cross section for spin-0 and spin-2 resonance masses between 0.9 and 4.0\TeV. The analysis performed by the ATLAS Collaboration in the boosted topology is based on the \pp data collected in 2015--2018 and covers resonance masses from 1 to 3\TeV~\cite{aad2020reconstruction}. No significant excess
is observed in that search either, and 95\% \CL exclusion limits are set in the signal range of 1--3\TeV. A separate
ATLAS search on the same data set targets spin-0 resonances, decaying to \HH, with masses from 251\GeV to 1.6\TeV~\cite{Run2atlasheavybbtautau}. While this ATLAS search primarily targets the resolved topology, it covers the same signal mass range as the present analysis, that is, between 1 and 1.6\TeV. The data are also found to be compatible with the
background-only hypothesis, with the largest deviation at a mass of 1\TeV,
corresponding to local and global significances of 3.1 and 2.0 standard deviations,
respectively.\par}

This paper focuses on the fully hadronic (\FH) and the mixture of one leptonic and one hadronic decay (\SL) of the \tautau pair in resonant Higgs boson pair production, where one \PH boson decays into \bbbar and the other into \tautau. The individual hadronically-decaying tau leptons are denoted as \tauh, while $\Pell$ denotes an electron or muon. In addition to $\PX \to \HH \to \bbbar \tautau$, other resonant \HH processes that can contribute to the same final states are also included as signal, namely $\PX \to \HH \to \bbbar\PW\PW \to \bbbar\Pelldag\PGn\Pq\Pq$ and $\PX \to \HH \to \bbbar \PV\PV \to \bbbar \Pelldag\Pelldag\PGn\PGn$. Here, $\Pelldag$ denotes a charged lepton of any flavor ($e$, $\mu$, or $\tau$), \Pq represents a quark, and \PV denotes a massive SM vector boson (\PW or \PZ), as used throughout this paper. In the simulated signal samples of $\PH\to\tautau$, the \Pgt leptons decay inclusively according to their SM branching fractions, while for $\PH\to\PW\PW$ and $\PH\to\PZ\PZ$ decays, we restrict the V decays to the final states listed above, i.e.,\ $\PW\PW\to \Pelldag\PGn\Pelldag\PGn$, $\Pelldag\PGn\Pq\Pq$, and $\PZ\PZ\to \Pelldag\Pelldag\PGn\PGn$. Therefore, the corresponding SM branching fractions~\cite{deFlorian:2016spz, PDG2024} of \PW and \PZ bosons to these final states are taken into account when calculating the signal yields. 

This analysis assumes a narrow resonance width, \ie, the natural width of the resonance (1\MeV) is significantly smaller than the experimental resolution (around 10\% of the mass of the resonance). The two bottom quarks from the Higgs boson decay are produced with a small angular separation and often merge into a single jet after hadronization. These jets are identified with a graph convolutional neural network called the \textsc{ParticleNet} jet tagger~\cite{qu2020jet}. 
In this Lorentz-boosted regime, the decay products of the \PGt lepton pair, whether \SL or \FH, tend to overlap spatially, making di-\PGt reconstruction particularly challenging. The reconstruction of such boosted \PGt leptons was first addressed in a similar CMS search using 8\TeV \pp collision data \cite{CMS:2015fjg}. 
In this analysis, we enhance the sensitivity of these methods by employing a dedicated \textsc{DeepTau}~\cite{CMS:2022prd} based tagger, 
optimized specifically for the reconstruction of boosted \PGt leptons. The SM processes with the largest contributions to the background are \ttjets (top quark-antiquark pair production), \wjets, and \zjets production, although the precise composition depends on the specific channel under consideration.

The structure of this paper is as follows: Section \ref{sec:cms-detector} provides an overview of the CMS detector, while Section \ref{sec:samples} describes the CMS data set and the simulation of physics processes. Section \ref{sec:eventReconstruction} details the object and event reconstruction, followed by Section \ref{sec:Selection}, which outlines the event selection and categorization. 
The methods used for background modeling are discussed in Section \ref{sec:backgroundEstimation}, and systematic uncertainties are addressed in Section \ref{sec:systematics}. Finally, the results are presented in Section \ref{sec:results}, and the paper is summarized in Section \ref{sec:summary}. Tabulated results for this analysis are provided in HEPData~\cite{hepdata}.

\section{The CMS detector}
\label{sec:cms-detector}
The central feature of the CMS apparatus is a superconducting solenoid of 6\unit{m} internal diameter, providing a magnetic field of 3.8\unit{T}. Within the solenoid volume are a silicon pixel and strip tracker, a lead tungstate crystal electromagnetic calorimeter (ECAL), and a brass and scintillator hadron calorimeter (HCAL), each composed of a barrel and two endcap sections. Forward calorimeters extend the pseudorapidity ($\eta$) 
coverage provided by the barrel and endcap detectors. Muons are detected in gas ionization chambers interleaved with the layers of the steel flux-return yoke outside the solenoid. A more detailed description of the CMS detector, together with a definition of the coordinate system used and the relevant kinematic variables, can be found in Refs.~\cite{Chatrchyan:2008zzk} and~\cite{CMS:2023gfb}.

Events of interest are selected using a two-tiered trigger system~\cite{CMS:2016ngn,CMS:2020cmk,CMS:2024aqx}. The first level, composed of custom hardware processors, uses information from the calorimeters and muon detectors to select events at a rate of around 100\unit{kHz} within a time interval of less than 4\mus. The second level, known as the high-level trigger, consists of a farm of processors running a version of the full event reconstruction 
software optimized for fast processing, and reduces the event rate to a few kHz before data storage.

\section{Data and simulated samples}
\label{sec:samples}

{\tolerance=800 This search utilizes data recorded by the CMS detector from 2016 to 2018, corresponding to an integrated luminosity of 138\fbinv. Events are selected using a set of triggers that require either missing transverse momentum (\ptmiss) or missing hadronic transverse momentum (\mht) 
to exceed 110\GeV, often in conjunction with additional criteria, such as the presence of a jet with $\pt > 80\GeV$. The primary vertex (PV)~\cite{CMS-TDR-15-02} is defined as the vertex with the largest $\pt^{2}$ sum of the physics objects originating from that vertex. Here, the physics objects are the charged leptons, the AK4 jets, and the \ptmiss. The missing transverse momentum vector $\ptvecmiss$ is defined as the negative vector sum of the momentum of all reconstructed particle candidates associated with the PV, projected onto the plane perpendicular to the beam direction. The missing transverse momentum \ptmiss is defined as the magnitude of $\ptvecmiss$. The observable \mht is defined as the magnitude of the vectorial
 \pt sum of all jets with $\pt>30\GeV$ and $\abs{\eta}<3.0$. The particle and jet reconstruction are described in detail in Section~\ref{sec:eventReconstruction}.\par}

{\tolerance=1000 Monte Carlo (MC) simulated samples of the main SM backgrounds and of the signal processes are used to optimize the event selection and to improve the background estimation. Simulated events of \ttjets production and single top quark processes ($t$-, $s$-, and $\PQt\PW$-channels) are generated at next-to-leading order (NLO) in quantum chromodynamics (QCD) using the {\POWHEG v1 and v2}~\cite{Frixione:2007vw,Alioli:2010xd} 
event generators. For the \ttjets process, the top quark \pt distribution is reweighted to match next-to-next-to-leading order (NNLO) theoretical computations~\cite{Czakon:2017wor}. Samples of \zjets and \wjets are generated at leading order (LO) using \MGvATNLO v2.6.5~\cite{Alwall:2014hca} with the MLM prescription~\cite{Mangano:2006rw} for matching jets from the matrix element calculation to the parton shower. Dedicated electroweak~\cite{Denner:2009gj, Denner:2011vu, Denner:2012ts, Kuhn:2005gv, Kallweit:2014xda, Kallweit:2015dum} and QCD (calculated with \MGvATNLO) NLO/LO $k$-factors, parameterized as functions of the generated boson \pt, are applied to \zjets and \wjets events. The QCD multijet events are simulated at LO either using the \PYTHIA~\cite{Sjostrand:2007gs} or \MGvATNLO event generators, with MLM matching~\cite{Alwall:2007fs} for the latter. Other processes with smaller contributions, such as diboson production, are generated at NLO using either \MGvATNLO with FxFx~\cite{Frederix:2012ps} matching or {\POWHEG v2}. All background samples are normalized using the most precise cross section evaluations available, which generally are calculated at NLO or NNLO.\par}

{\tolerance=800 Signal samples of narrow-width $\Gamma_{\PX} = 1\MeV$ spin-0 and spin-2 resonances, produced via gluon fusion and decaying to \HH, are simulated at LO using {\MGvATNLO}. The spin-0 (spin-2) signal is modeled as a radion-like scalar (graviton-like) resonance, with spin correlations handled through the corresponding matrix elements. Resonance masses range from 1 to 4.5\TeV. For spin-0 resonances, mass points are spaced by 200\GeV between 1 and 2\TeV, and by 500\GeV above 2\TeV. For spin-2 resonances, the spacing is 250\GeV between 1 and 2\TeV, and 500\GeV above 2\TeV.\par}

{\tolerance=1000 The initial-state partons are modeled with the NNPDF 3.1 NNLO~\cite{Ball_2017} parton distribution function (PDF) set. Parton showering and hadronization are handled by {\PYTHIA v8.230}~\cite{Sjostrand:2007gs} using the \textsc{CP5} tune~\cite{CMS:2019csb} for most samples. All signal and background samples are processed using \GEANTfour~\cite{Agostinelli:2002hh} 
to provide a full simulation of the CMS detector, including simulation of the triggers. The effects of additional \pp interactions in the same or adjacent bunch crossings, referred to as pileup, are included in all simulated samples. To match the simulated distribution of pileup interactions with the one observed in data, a reweighting procedure is implemented. Correction factors are derived and applied to the simulated 
samples to match the trigger efficiencies measured in data. Additional corrections are applied to cover residual differences between data and simulation that arise from the lepton identification and reconstruction efficiencies, as well as from identification efficiencies for jets identified to originate from \PQb quarks.\par}

\section{Event reconstruction}
\label{sec:eventReconstruction}
A particle-flow (PF) algorithm~\cite{Sirunyan:2017ulk} reconstructs and identifies each individual particle through an optimized combination of information from the various elements of the CMS detector.
The energy of each electron is determined from a combination of the electron momentum as measured by the tracker, the energy of the corresponding ECAL cluster, and the energy sum of all bremsstrahlung photons spatially compatible with originating from the electron track. The energies of muons are obtained from the curvature of the corresponding track.
The energies of charged hadrons are determined from a combination of their momenta measured in the tracker and the matching ECAL and
 HCAL energy deposits, corrected for zero-suppression effects and for the response function of the calorimeters to hadronic showers. 
Finally, the energies of neutral hadrons are obtained from the corresponding corrected ECAL and HCAL energies.

The identified particles are clustered into jets using the anti-\kt algorithm~\cite{Cacciari:2008gp}, implemented in the {\FASTJET} package~\cite{Cacciari:2011ma}. Two distance parameters are used in the analysis, 0.4 and 0.8, yielding jet collections referred to as AK4 and AK8 jets, respectively. The PF candidates and AK4 jets are used in the definitions of \ptmiss and \mht. 
The AK4 jets are used primarily to reject or select events with top quarks, while the larger AK8 jets are used to identify and contain hadronically decaying \PH boson candidates.

{\tolerance=1000 Additional selection criteria are applied to each event to remove spurious jet-like features originating from isolated noise patterns in certain HCAL regions~\cite{CMS:2019ctu}. The AK4 and AK8 jets must have $\pt>20\GeV$ and $\pt>200\GeV$, respectively and $\abs{\eta}<2.4$, to be considered in the subsequent steps of the analysis.\par}

To improve the stability of the jet mass and substructure variables in events with high pileup, the pileup-per-particle identification (PUPPI) algorithm~\cite{Bertolini2014} is applied to the AK8 jets used to identify hadronic boson decays. The PUPPI algorithm uses the local distribution of particles, event pileup properties, and tracking information to compute a weight describing the likelihood for each particle to originate from a pileup interaction. The weight is used to rescale the particle four-momentum, superseding the need for further jet-based pileup corrections.

Large AK8 jets originating from the dominant \bbbar decays of \PH bosons are likely to have two displaced vertices because of the long lifetime and large mass of the \PQb quarks. A graph convolutional neural network, \textsc{ParticleNet}~\cite{qu2020jet}, is used to distinguish \PH boson decays into \bbbar from background jets by leveraging the properties of jet PF constituents as input features. As with all heavy-flavor jet classifiers, displaced tracks 
and secondary vertices are among the most important features. The multiclass \textsc{ParticleNet} algorithm provides several output scores, each ranging from 0 to 1, representing the probability that a jet originates from a specific process, such as \PH boson decay to \PQb quarks ($P(\PH\to\bbbar)$), or a light-flavor quark or gluon ($P(\text{QCD})$). Following the standard approach for constructing a binary discriminant to distinguish \PH boson jets from the dominant QCD multijet background~\cite{CMS:2025kje}, the \textsc{ParticleNet} score is defined as $P(\PH\to\bbbar) / ( P(\PH\to\bbbar) + P(\text{QCD}))$. In addition to classification, \textsc{ParticleNet} performs a regression task to estimate the mass of the AK8 jet, referred to in this work as the \textsc{ParticleNet} regressed mass \MHbb. In this analysis, we define a signal-enriched region (SR), defined as $100 < \MHbb < 150\GeV$, and a second region, called the sideband (SB), defined as
$\MHbb < 100\GeV$ or $\MHbb > 150\GeV$, enriched in background processes.

The AK4 jets originating from \PQb quarks are identified using the \textsc{DeepJet} algorithm~\cite{bols2020jet}, a deep neural network that combines secondary vertex information, track-based variables, and PF jet constituents (neutral and charged-particle candidates). Events are vetoed if they contain at least one \PQb-tagged AK4 jet that does not overlap with the selected AK8 jet or the leptons in the event, to further suppress the \ttbar background.

A series of dedicated algorithms are used to reconstruct and identify \PH bosons decaying to \tautau pairs with either \FH or \SL signatures. These procedures are described in the rest of this section.

{\tolerance=800 Highly collimated \FH or \SL pairs are quite challenging to reconstruct and require the use of a special boosted reconstruction algorithm~\cite{CMS:2018jrd}. The procedure begins by using the Cambridge--Aachen algorithm~\cite{Wobisch:1998wt} with a distance parameter of 0.8 to identify jets with a large cone size, called CA8 jets.
For each CA8 jet with $\pt>100\GeV$, the last step of the clustering is retracted, resulting in two subjets. 
If these subjets are found to have $\pt>10\GeV$ and satisfy the mass drop condition, which requires $\max(\msjone, \msjtwo)/\mcajet<2/3$, the two subjets are used as inputs to the standard $\Pgt$ lepton reconstruction, the ``hadron-plus-strips" (HPS) algorithm~\cite{Chatrchyan:2012zz}, to build the \tauh candidates. 
Here, \msjone and \msjtwo are the masses of the two subjets, and \mcajet is the mass of the original CA8 jet. 
If the \pt and mass drop conditions are not met, the unclustering and identification procedures are repeated iteratively for the most energetic subjet. 
We will refer to \PGt leptons reconstructed in this fashion as boosted \tauh.\par}

The boosted reconstruction algorithm is not well suited for \PH bosons with lower Lorentz boost, as is typical of lower resonance mass signal hypotheses around 1 to 1.5\TeV. To cover this topology, we instead reconstruct single \tauh candidates using AK4 jets as inputs to the same HPS algorithm. These are referred to as standard \tauh.

The final selection of \tauh candidates, whether boosted or standard, requires $\abs{\eta} < 2.5$ and $\pt > 20\GeV$, along with a convolutional neural network based tagger score above a specified threshold corresponding to per-\tauh signal efficiency greater than 95\%. The identification of boosted \tauh candidates was developed specifically for this analysis and is based on the \textsc{DeepTau} architecture originally designed for standard \PGt leptons. This tagger, 
referred to as \textsc{BoostedDeepTau} throughout the article, is trained to distinguish genuine boosted \PGt leptons from background jets by leveraging both high-level features derived from the boosted \PGt reconstruction and low-level detector information. Among its important inputs are a global event-level variable related to the average energy deposition in the event and 42 high-level variables that capture the kinematic properties, isolation, and finite lifetime signature of boosted \PGt leptons. Many of these variables have also proven effective in previously developed \PGt identification algorithms, such as a discriminator based on a boosted decision tree (MVA Iso)~\cite{CMS:2018jrd}.

The training of the \textsc{BoostedDeepTau} algorithm is performed using simulated samples. Genuine boosted \PGt leptons are taken from simulated spin-0 signal samples, while the dominant backgrounds---QCD multijet, \ttbar, and Drell--Yan processes---are included to ensure comprehensive coverage of the relevant phase space. The training samples include various mass points and are shuffled to balance the boosted \PGt lepton population across $\pt$, $\eta$, and other relevant variables. 
The resultant \textsc{BoostedDeepTau} tagger shows a discrimination power that is a factor 2--4 better than MVA Iso evaluated on boosted \PGt leptons for $\pt<100\GeV$ and more than a factor of 10 better for $\pt>100\GeV$. Additional information about the \textsc{BoostedDeepTau} algorithm can be found in Ref.~\cite{CMS-DP-2025-047}. 

Electrons from leptonic tau decays are reconstructed in the region $\abs{\eta}<2.5$ by matching energy deposits in the ECAL with tracks reconstructed in the tracker~\cite{Khachatryan:2015hwa}. Electron identification is based on the distribution of energy deposited along the electron trajectory, and the direction and momentum of the track in the inner tracker. Additional requirements are applied to remove electrons produced through photon conversions. Electrons are also required to be isolated from other particles in the detector, by imposing an upper threshold on the relative isolation parameter. The electron isolation parameter is defined as the magnitude of the \pt sum of all the PF candidates (excluding the electron) within $\DR< 0.3$ around the electron direction, after the contributions from pileup and particles associated with reconstructed \tauh candidates within the isolation cone have been removed. The resulting sum is then divided by the electron \pt, yielding a corrected relative isolation variable $I_\text{rel}^\Pe$, which is required to satisfy $I_\text{rel}^\Pe < 0.112 + 0.506/\pt$ in the ECAL barrel ($\abs{\eta}<1.479$) and $I_\text{rel}^\Pe < 0.108 + 0.963/\pt$ in the ECAL endcaps ($1.479<\abs{\eta}< 2.5$), with \pt in \GeV.

Muons from leptonic tau decays are reconstructed within the acceptance of the CMS muon system, $\abs{\eta}<2.4$, using information from both the muon spectrometer and the silicon tracker~\cite{Chatrchyan:2012xi}. Muon candidates are identified based on the compatibility of tracks reconstructed in the silicon tracker with tracks reconstructed from a combination of hits in both the tracker and the muon detector. In addition, the trajectory is required to be compatible with originating from the primary vertex, and to have a sufficient number of hits in the tracker and muon systems. Muons are required to be isolated by applying a relative isolation criterion. Specifically, the scalar sum of the \pt of all PF candidates within a cone of $\DR < 0.4$ around the muon direction---excluding the muon itself---is computed. Contributions from particles associated with reconstructed \tauh candidates within the isolation cone are subtracted. The resulting sum is then divided by the muon \pt, yielding a corrected relative isolation variable $I_\text{rel}^\PGm$, which is required to satisfy $I_\text{rel}^\PGm < 0.25$. This procedure is very similar to that followed for the electrons.

 The standard (boosted) \tauh candidates are either combined with another standard (boosted) \tauh candidate or with a leptonic \PGt decay identified as an electron or a muon, to form \tautau pairs. Both standard \tauh and boosted \tauh candidates are required to be distinct from the selected AK8 jet in the event.

\section{Event selection}
\label{sec:Selection}
As mentioned earlier in Section~\ref{sec:samples}, events are selected using a suite of \ptmiss and \mht triggers. For the heavy parent resonance masses considered in this analysis, the H bosons are produced with large Lorentz boosts. As a result, the neutrinos from the \tautau decay carry substantial momentum, and signal events are therefore expected to exhibit large \ptmiss, ensuring high efficiency of the \ptmiss and \mht triggers. A stringent offline requirement of $\ptmiss>180\GeV$ is applied after the full event reconstruction, to ensure a stable trigger efficiency and to suppress the background contribution from multijet events. The efficiency of the trigger for events satisfying the offline event selection, measured in an independent sample of events selected with muon triggers, is found to be greater than 95\%, with an uncertainty of 1.1\%.

All events are required to contain one \PH boson candidate decaying to either $\Pell\tauh$ or $\tauh\tauh$. If multiple candidate pairs are possible, the pair with the highest visible \pt is selected. The other \PH boson candidate decaying to \bbbar is reconstructed as an AK8 jet that does not overlap with the \tautau system. The same AK8 jet kinematic selection and identification requirements are applied in both channels. If multiple AK8 jet candidates are present, the jet with the highest \pt is selected. The $\PH \to \bbbar$ candidate jet is also required to satisfy the \textsc{ParticleNet} \bbbar tagging requirement corresponding to a working point with 80\% 
efficiency for selecting signal jets from $\PH \to \bbbar$ decays. This criterion significantly suppresses background from QCD multijet events and jets produced in association with vector bosons. Events with top quark pairs or single top quarks are suppressed by vetoing events with \PQb-tagged AK4 jets that do not overlap with the AK8 $\PH\to\bbbar$ candidate and are distinct from the identified $\PH\to\tautau$ objects.

{\tolerance=1000 Additional selection criteria are applied to suppress SM contributions, including those from meson and baryon resonances, \wjets, \ttbar, and single top quark production. To reduce contamination from misidentified \PGt leptons in \wjets events, the angular separation $\DR_{\tautau}$ between the two \PGt candidates is required to be less than 1.5. Additionally, we require the difference in the azimuthal angle between the AK8 jet, identified as the \PH boson through the $\PH\to\bbbar$ decay, and \ptvecmiss to be greater than 1. This requirement ensures that the Higgs boson candidate and the missing transverse momentum are well separated in the transverse plane, typically appearing on opposite sides of the event, as expected in signal-like topologies. To reduce contributions from low-\pt radiation, the soft-drop (SD)~\cite{Larkoski:2014wba} jet mass of the $\PH\to\bbbar$ candidate jet is required to be greater than 30\GeV. Furthermore, the visible mass of the \PH boson decaying into \tautau must be greater than 20\GeV to suppress low-mass and quarkonium backgrounds.\par}

An algorithm called \textsc{FastMTT}~\cite{Matyszkiewicz:2025gal}, based on a dynamical likelihood approach, is used to reconstruct the four momentum of the 
di-\PGt system, including the missing information from neutrinos. This method incorporates the measured momentum of the visible \PGt decay products, the reconstructed \ptmiss, and its resolution to enhance the reconstruction accuracy. To speed up the computation, the algorithm applies the collinear approximation~\cite{Ellis:1987xu}, which assumes that the neutrinos 
from \PGt decays are collinear with the visible decay products. The $\PH \to \bbbar$ candidate four-momentum is formed using kinematic variables obtained from the standard AK8 jet reconstruction algorithm, to which energy and resolution corrections are applied to match the response in data. 
\begin{figure}[tbp]
\centering
    \includegraphics[width=0.48\textwidth]{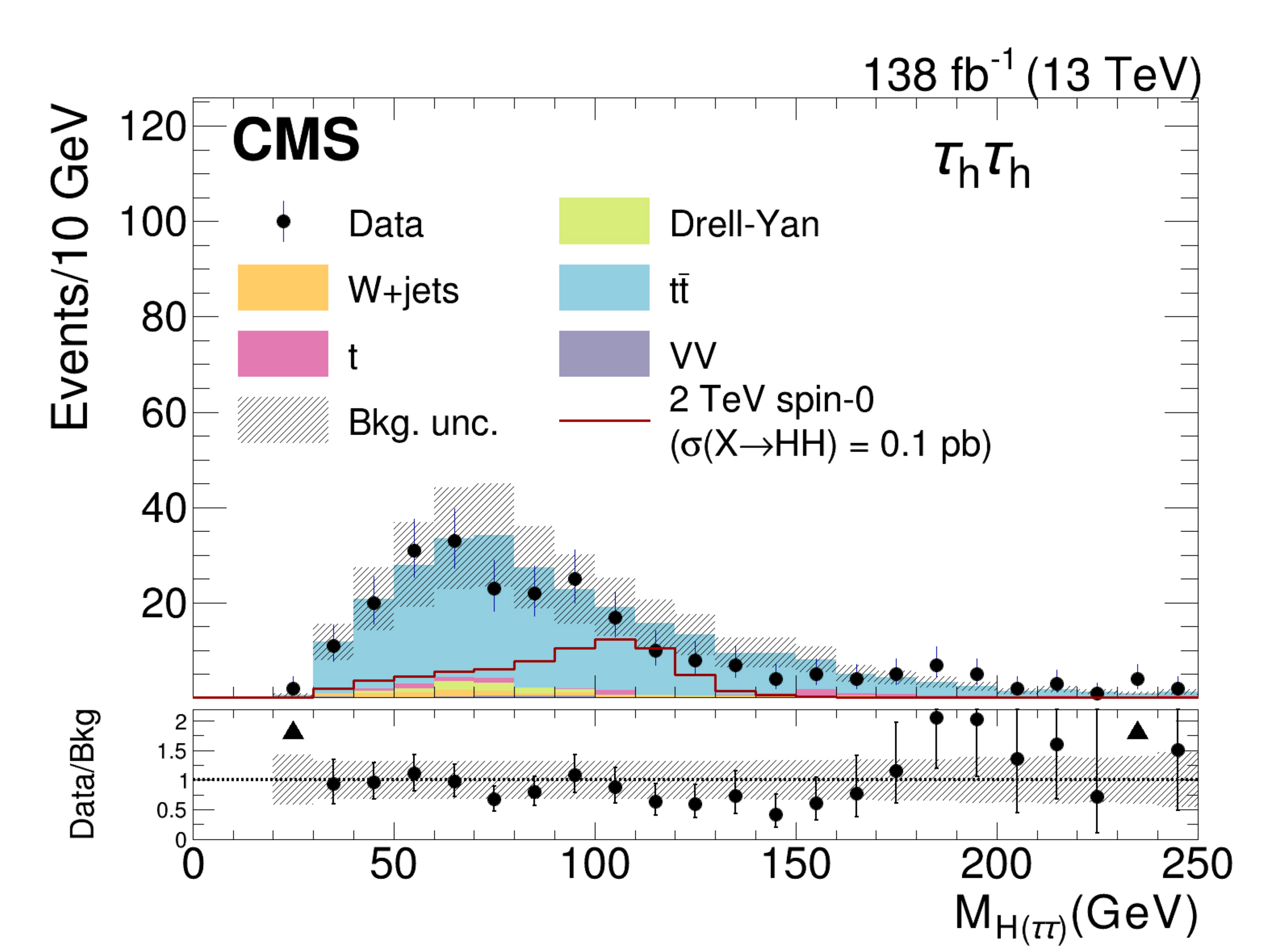}
    \includegraphics[width=0.48\textwidth]{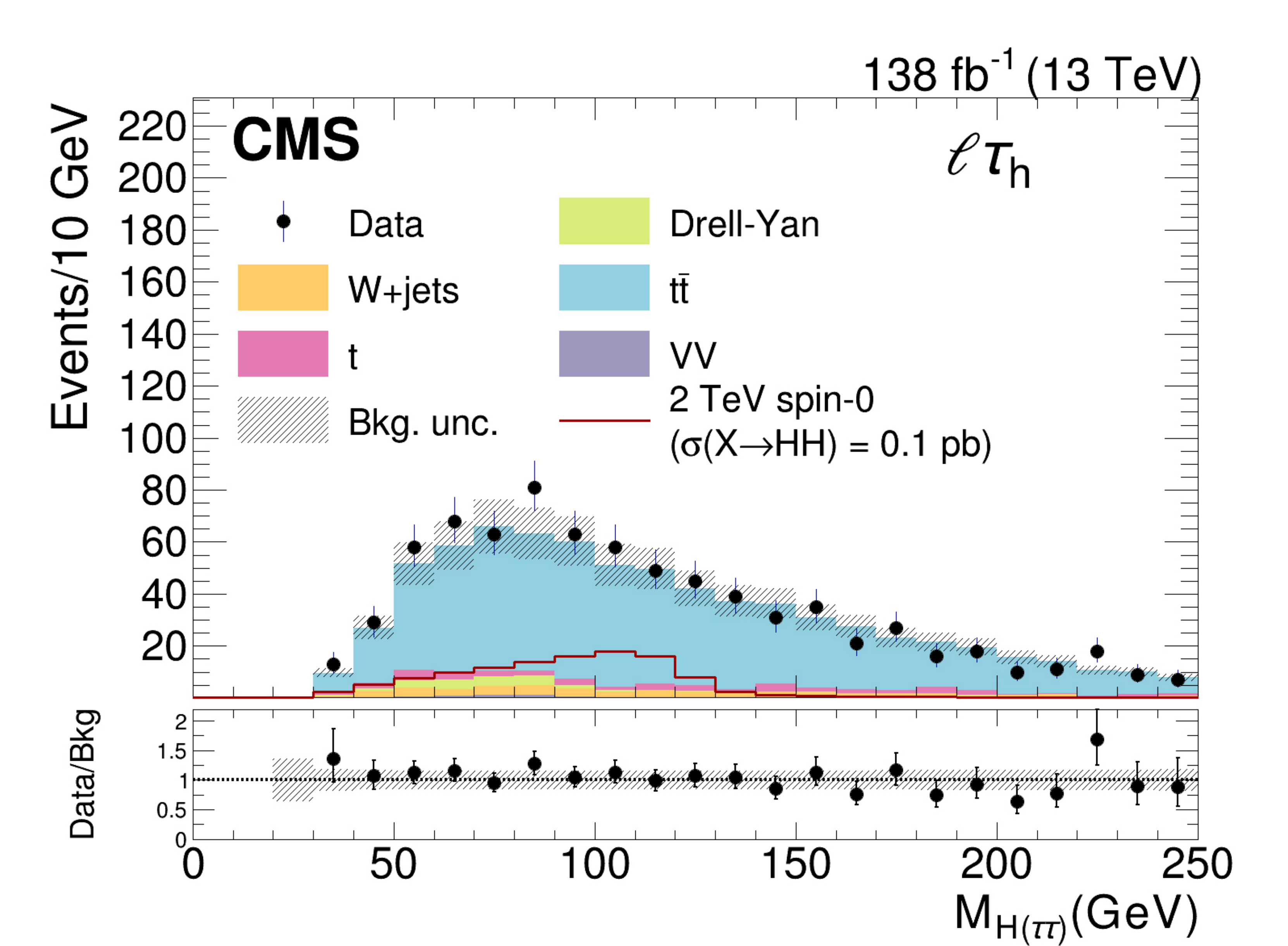}
\caption{Distribution of the invariant mass of the di-\PGt system, reconstructed with the \textsc{FastMTT} algorithm, after the full event selection in the \FH (\cmsLeft) and \SL (\cmsRight) channels. The data (solid circles) are compared to the background simulation (filled histograms), where the gray bands represent the total background uncertainty, obtained from the post-fit values of the dominant systematic uncertainties and the statistical uncertainties in the simulated samples. The \(\mathrm{X}\to\mathrm{HH}\) signal simulation (solid red line) is overlaid and normalized to \(\sigma(\mathrm{X}\to\mathrm{HH})=0.1~\unit{pb}\) for illustration. The ratio between the data and the total expected background contribution is shown in the lower panel, where a solid black triangle indicates those bins where the ratio exceeds the axis range.
\label{fig:bkgest:fastmttdist}
}
\end{figure}

\begin{figure}[tbp]
\centering
    \includegraphics[width=0.48\textwidth]{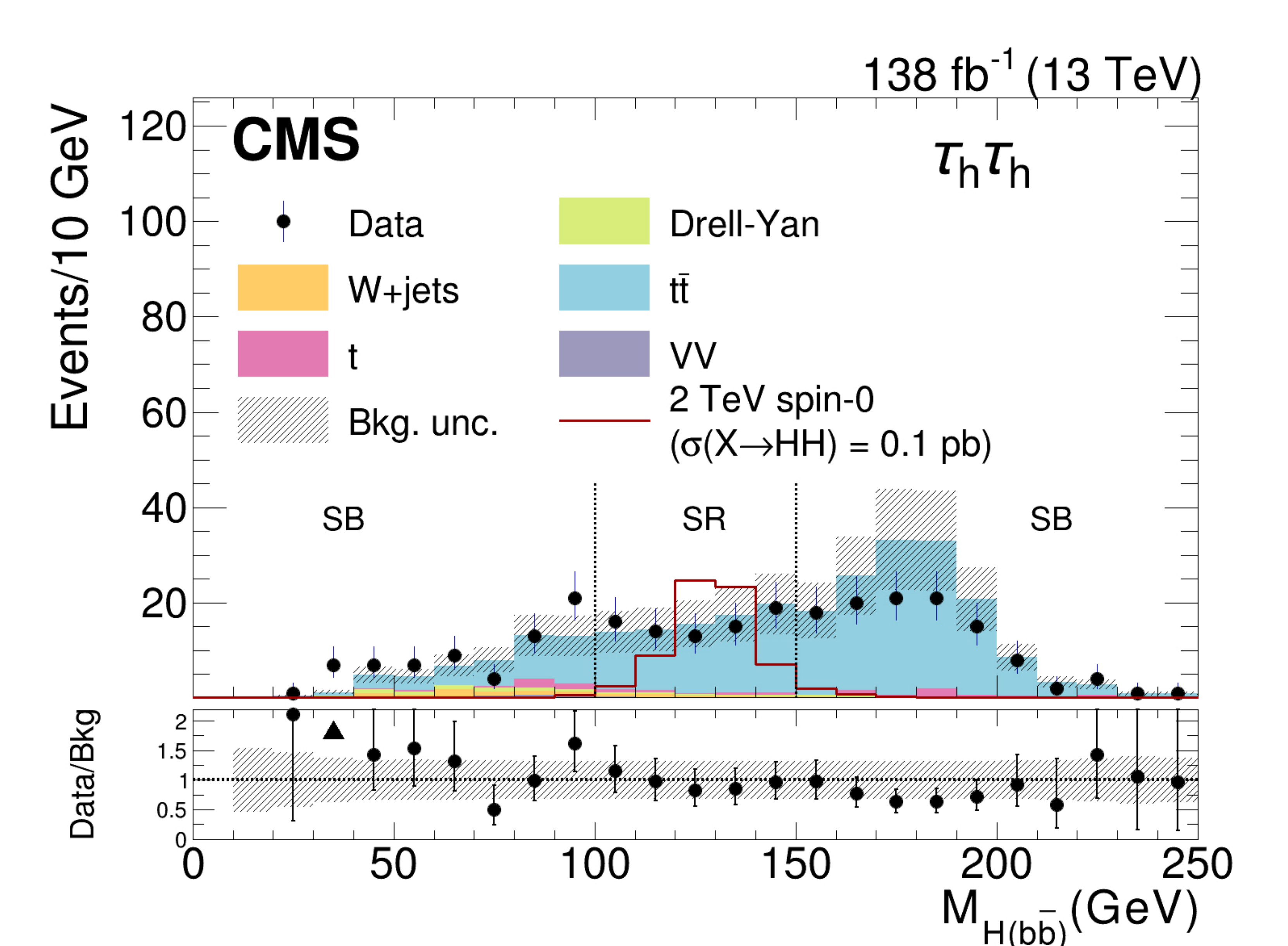}
    \includegraphics[width=0.48\textwidth]{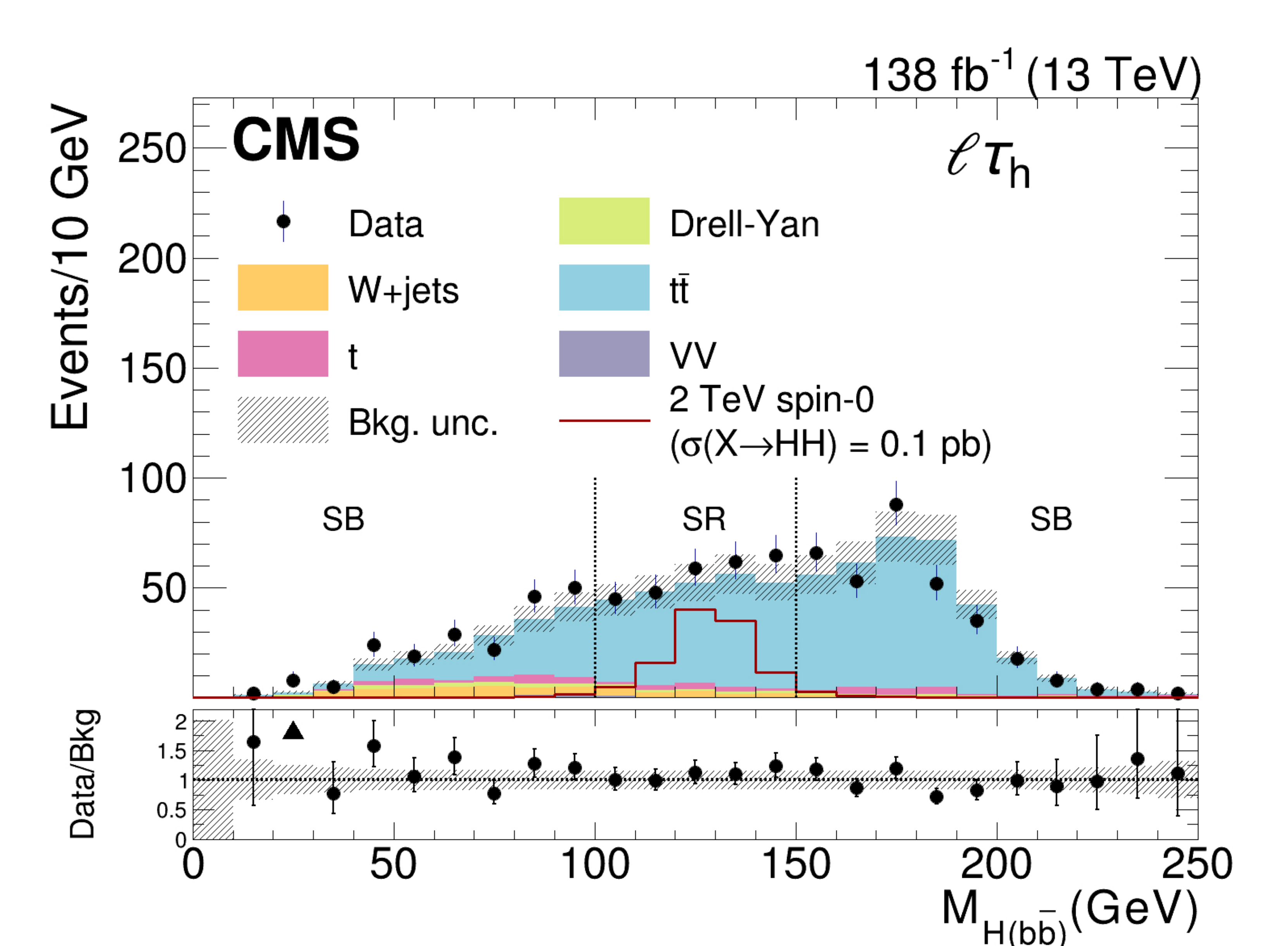}
\caption{Distribution of \MHbb obtained from the leading AK8 jet in the event after the full event selection for the \FH (\cmsLeft) and \SL (\cmsRight) channels. 
The signal-enriched region (SR) is defined as $100 < \MHbb < 150\GeV$. The sideband (SB) is immediately adjacent to the SR, on either side. 
The data (solid circles) are compared to the background simulation (filled histograms), where the gray bands represent the total background uncertainty, obtained from the post-fit values of the dominant systematic uncertainties and the statistical uncertainties in the simulated samples. The \(\mathrm{X}\to\mathrm{HH}\) signal simulation (solid red line) is overlaid and normalized to \(\sigma(\mathrm{X}\to\mathrm{HH})=0.1~\unit{pb}\) for illustration. The ratio between the data and the total expected background contribution is shown in the lower panel, where a solid black triangle indicates those bins where the ratio exceeds the axis range.   
\label{fig:bkgest:regiondefintion}
}
\end{figure}

The reconstructed resonance candidate mass $M_{\PX}$, defined as the invariant mass of the tagged $\PH \to \bbbar$ candidate jet and the \textsc{FastMTT} reconstructed di-\PGt system, is required to be larger than 750\GeV to ensure full trigger and reconstruction efficiencies. All selections, except for the \textsc{ParticleNet} \bbbar-tagging and the AK4 \PQb-tagged event veto, are collectively referred to as the ``pre-selection'' in this analysis. The distributions of important kinematic variables for the full event selection are shown in Figs.~\ref{fig:bkgest:fastmttdist} and \ref{fig:bkgest:regiondefintion}. 

The full event selection efficiency, including the trigger selections, for spin-0 (spin-2) resonances ranges from 2\% (2.7\%) to 13.5\% (14.4\%) in the $\tauh\tauh$ channel and from 4\% (5.4\%) to 11\% (11.8\%) in the $\Pell\tauh$ channel, depending on the mass of the resonance. Additionally, events from the $\HH \to \bbbar\PV\PV$ processes contribute approximately 20\% of the total yield of the $\HH \to \bbbar\tautau$ signal. Around 90\% of the 
selected $\HH \to \bbbar\PV\PV$ events originate from genuine $\Pell\tauh$ or $\tauh\tauh$ final states.
To extract a potential resonant \HH signal, a maximum likelihood fit is performed using the $M_{\PX}$ distributions in data in the \SL and \FH channels. These fits are performed using the CMS statistical analysis tool \textsc{Combine}~\cite{hayrapetyan2024cms}, which is based on the \textsc{RooFit}~\cite{Verkerke:2003ir} and \textsc{RooStats}~\cite{Moneta:2010pm} frameworks. Accurate modeling of the major backgrounds is essential for the signal extraction. 
The strategy for estimating this is described in the following section.

\section{Background estimation}
\label{sec:backgroundEstimation}
After the event pre-selection, the leading SM background contributions in the different regions are from \ttbar, \zjets, and \wjets production.  
Following the full event selection, the \ttbar process becomes the dominant background, as shown in Figs.~\ref{fig:bkgest:fastmttdist} and~\ref{fig:bkgest:regiondefintion}.
The QCD multijet contribution is negligible owing to the lepton requirement in the \SL channel, and the combined effect of the 
\textsc{BoostedDeepTau}/\textsc{DeepTau} and the \textsc{ParticleNet}-based AK8 $\PH\to\bbbar$ selections in the \FH channel.  

The SR and SB regions are defined using the reconstructed \MHbb distribution, as shown in Fig.~\ref{fig:bkgest:regiondefintion}. The \MHbb distribution is used only to categorize events into regions, whereas the final discriminant used in the fit is the $M_{\PX}$ distribution. We use the data in the SB to constrain the normalization and the $M_{\PX}$ shape of the top quark background in the SR, where the top quark background in both the SR and the SB combines the contributions from \ttbar and single top quark production. This is achieved by introducing unconstrained, bin-by-bin multiplicative parameters for each bin of the $M_{\PX}$ distribution. These parameters simultaneously adjust the yield of the top quark background per $M_{\PX}$ bin in both the SR and SB for the corresponding channel during the fit.

\section{Systematic uncertainties}
\label{sec:systematics}
\begin{table*}[htbp]
  \centering
  \topcaption{Summary of the dominant systematic uncertainty sources and the typical size of their inferred variation in the resonance yield. Unless otherwise specified, the impact is the same for the \FH and \SL channels. Sources with an overall impact of less than $1\%$ on the resonance yield, such as electron or muon reconstruction and identification, are not listed in the table.}
  \begin{tabular}{lc}
    Systematic source & Variation \\ 
    \hline
    Boosted \tauh lepton identification & 20\% for \FH, 10\% for \SL \\
    \tauh energy scale                  & 10\% \\
    Renormalization and factorization scales        & 5--10\% \\
    PDF uncertainties                               & 5--7\% \\
    \textsc{ParticleNet} AK8 \bbbar tagging               & 5\% (signal jets) \\
    Jet energy scale and resolution                 & 2--5\% \\
    Pileup modelling                                & 1--5\% \\
    Electroweak and QCD $k$-factors for $\mathrm{V}+$jets    & 1--5\% \\
    Integrated luminosity                           & 1.6\% \\
    \ptmiss trigger          & 1.1\% \\
  \end{tabular}
  \label{tab:systematics}
\end{table*}

Various sources of systematic uncertainties are taken into account, in most cases affecting both the signal and the background.
We distinguish between two types of systematic uncertainties: those that affect only the overall normalization of a process and those that also impact the shape of the reconstructed resonance mass spectrum, which serves as the primary fit observable in the analysis (Table~\ref{tab:systematics}).

The following sources of uncertainty are taken into account using constrained nuisance parameters in the fit for all processes. These parameters are used to encode the impact of systematic uncertainties on the expected signal and background yields and shapes. Unless specified otherwise, these are considered correlated across the data taking years. An exception is made for the top quark background. The constrained nuisance parameters included for the top quark background account only for the statistical uncertainties. Systematic uncertainties that impact the top quark background are included in the simultaneous fit to data in the SR and SB through unconstrained bin-by-bin multiplicative parameters, as described in Section~\ref{sec:backgroundEstimation}. The resulting post-fit uncertainties in these parameters reflect the fit uncertainty in the top quark background prediction.

\begin{itemize}
    
\item \textit{Integrated luminosity}: The integrated luminosities for the 2016, 2017, and 2018 data-taking years have 1.2--2.5\% individual uncertainties~\cite{CMS-LUM-17-003,CMS-PAS-LUM-20-001}, while the overall uncertainty for the 2016--2018 period is 1.6\%.

\item \textit{\ptmiss trigger}: A normalization uncertainty of 1.1\% is assigned to all processes, based on the level of agreement observed between data and simulation. 

\item \textit{Lepton identification and isolation}: A normalization uncertainty is assigned to account for the residual differences between data and simulation in the lepton identification and isolation efficiencies after applying all corrections. In the $\SL$ channel, a 0.4\% uncertainty is applied for both electron identification and isolation, and 1\% (0.1\%) for muon identification (isolation), affecting all processes.

\item \textsc{ParticleNet} \textit{AK8 \bbbar-tagging}: A 5\% normalization uncertainty is applied to signal AK8 jets to account for the residual data-to-simulation mismodeling of \PQb tagging. This uncertainty reflects the relatively uniform effect of the \PQb tagging uncertainty across the \pt spectrum of the AK8 jets.

\end{itemize}

The following sources of uncertainty affect the shape of the distributions, as well as the normalization of the various backgrounds and the signal, and are applied to all search channels.

\begin{itemize}
\item \textit{PDF uncertainties}: Uncertainties from the PDFs are estimated by reweighting the samples with the NNPDF3.0~\cite{Ball:2014uwa} replicas~\cite{Butterworth:2015oua} for the 2016 data set and NNPDF3.1~\cite{Ball_2017} for the 2017 and 2018 data sets. The variations across the replicas are used to construct an envelope that captures the spread in predictions, which is then taken as the systematic uncertainty. 

\item \textit{Factorization and renormalization scales}:
Variations in the renormalization and factorization scales by factors of 0.5 and 2.0 are considered independently with the exception of simultaneously varying both. Their impact is calculated from the envelope of these variations and included as systematic uncertainties for all processes. 

\item \textit{Pileup modeling}.
{\tolerance=1000 Systematic uncertainties from pileup modeling are taken into account by varying the total inelastic cross section used to calculate pileup distributions in simulation by $\pm4.6\%$~\cite{Sirunyan:2018nqx}.\par}

\item \textit{Jet energy scale and resolution}: Reconstructed jet four-momenta in the simulation are varied according to the uncertainties in the jet energy scale, which are split into 11 different sources. 
Additionally, the \pt of the jets is stochastically changed within the detector resolution. These uncertainties are coherently propagated to all observables, including \ptmiss~\cite{Khachatryan:2016kdb}. Some of the 11 sources are uncorrelated across the data taking years. The average effect from these sources of uncertainty ranges from 2 to 5\%.

\item \textit{\ptmiss mismodeling}: An additional source of uncertainty specific to \ptmiss arises from unclustered energy, primarily due to calorimeter deposits not associated with any reconstructed object. This unclustered energy is varied within its uncertainties, and the resulting impact is propagated to the \ptmiss.

\item \textit{\PQb tagging}: The \PQb tagging and light-flavor misidentification efficiency scale factors, along with their associated uncertainties, are measured using independent control samples~\cite{Sirunyan:2017ezt}. The uncertainties in these scale factors are then propagated to the analysis. Some sources of these uncertainties are uncorrelated across data taking years.  

\item \textit{Simulation sample size}: Uncertainties due to the limited size of the simulated signal and background samples are included by allowing each bin of the distributions used in the signal extraction to fluctuate independently according to the statistical uncertainties in simulation~\cite{BARLOW1993219}.

\item \textit{Uncertainty related to ECAL mistiming}: Partial mistiming of signals in the forward regions of the ECAL endcaps led to a minor reduction in trigger efficiency. Simulations are corrected to emulate the behavior of the data, and the uncertainty in these corrections is propagated to the distributions used in the signal extraction. 

\item \textit{Electroweak and QCD $k$-factors}: Uncertainties in the NLO/LO $k$-factors~\cite{Denner:2009gj, Denner:2011vu, Denner:2012ts, Kuhn:2005gv, Kallweit:2014xda, Kallweit:2015dum} calculated for \wjets and \zjets processes are considered. These uncertainties account for missing higher-order corrections. For QCD, this comes from variations of the factorization and renormalization scales. For electroweak processes, an estimate of the size of the missing higher-order corrections is obtained by taking the difference 
between applying and not applying the NLO/LO electroweak $k$ factors.

\item \textit{Hadronically decaying \PGt lepton identification}: Uncertainties in the \textsc{DeepTau} efficiency corrections, for the discrimination of genuine \tauh against jet, electron, and muon are correlated across \PGt decay modes and years for the standard \PGt leptons. Correction factors are derived to account for differences between data and simulation in the identification of boosted \PGt leptons from the \textsc{BoostedDeepTau} algorithm. 
These were derived using boosted \PZ boson decays, utilizing events with a spatially close muon--\PGt pair. A simultaneous fit is then performed in the $\PGm\tauh$ SR and a dimuon control region, with the latter constraining the normalization of the Drell--Yan process. The data-to-simulation factors were found to be 0.90 per identified boosted \PGt lepton, with an associated uncertainty of 10\%. 
This corresponds to a normalization uncertainty of 20\% in the \FH channel and 10\% in the \SL channel~\cite{CMS-DP-2025-047}.

\item \textit{Hadronically decaying \PGt lepton energy scale}: For the \tauh energy scale contribution to the systematic uncertainty, the four-momenta of reconstructed and identified \tauh, including both standard and boosted \tauh objects, are varied by $\pm 3\%$~\cite{CMS:2018jrd}. The resulting variations are propagated to all kinematic observables used in the analysis. For minor backgrounds, this effect is accounted for by a 10\% normalization uncertainty, while for the signal a dedicated shape systematic is assigned.
\end{itemize}

\section{Results}
\label{sec:results}

\begin{figure}[btp]
\centering
    \includegraphics[width=0.48\textwidth]{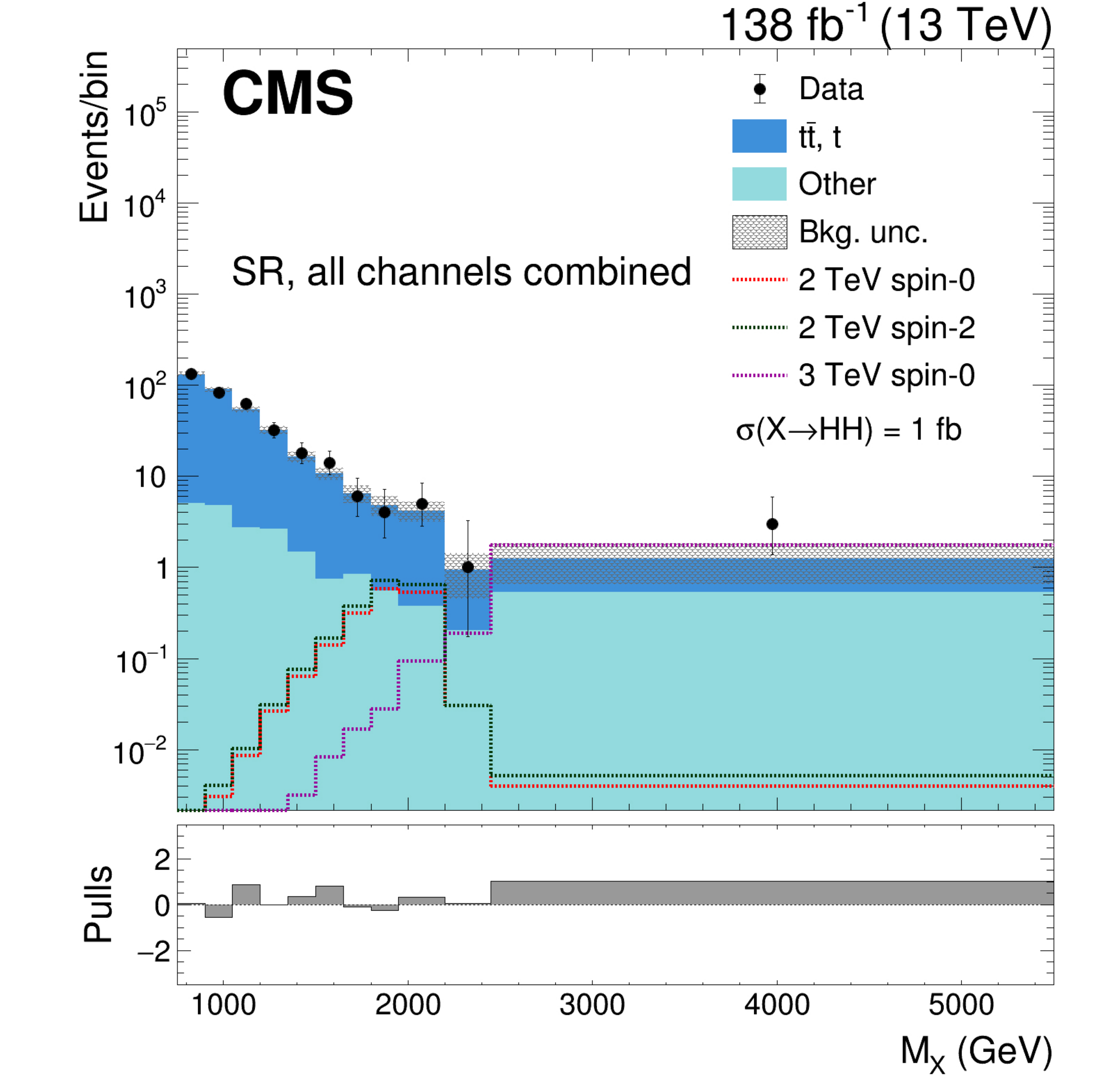}
    \includegraphics[width=0.48\textwidth]{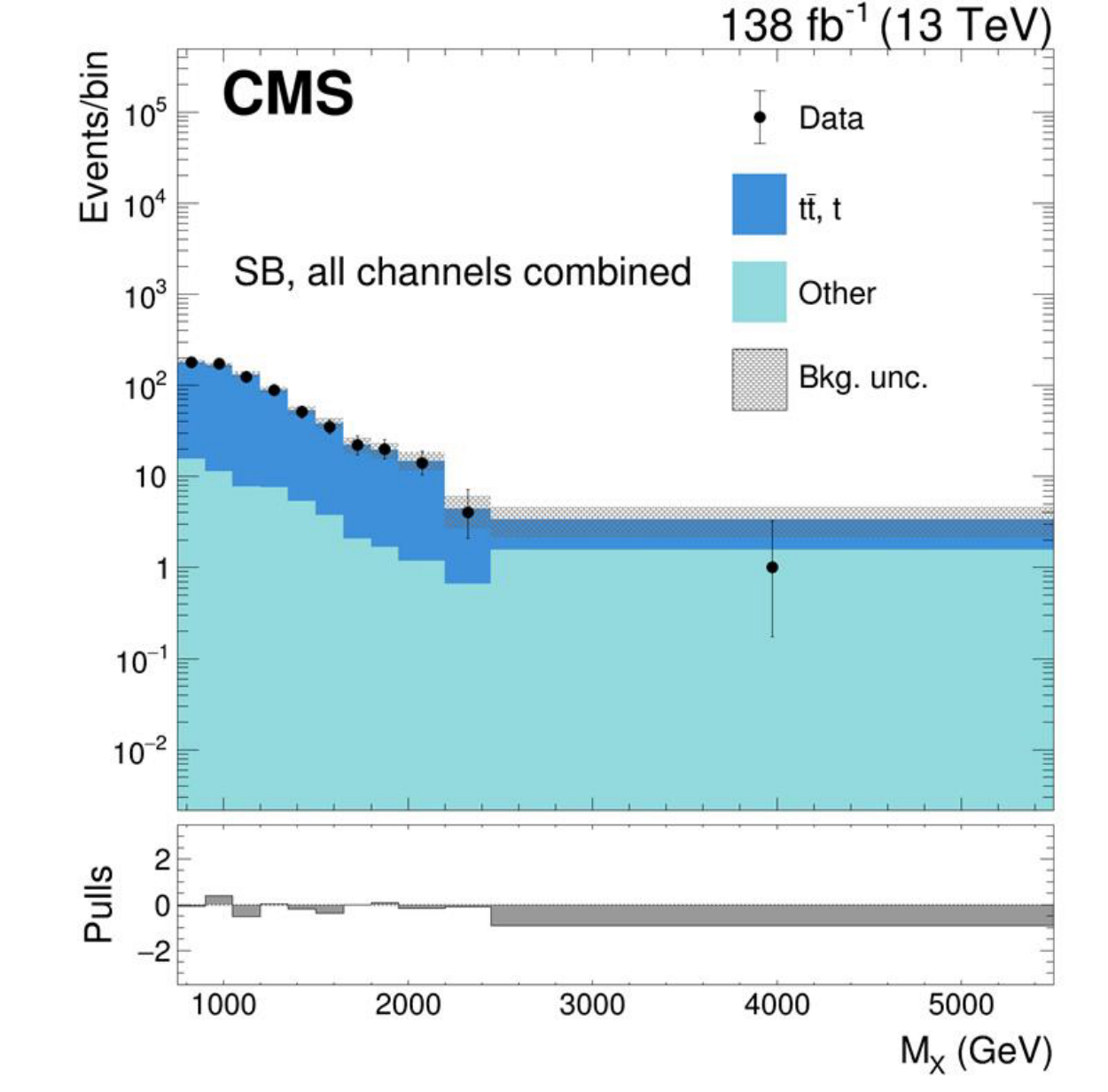}
    \caption{Post-fit reconstructed mass distribution of resonance \PX in the SR (\cmsLeft) and SB (\cmsRight) after applying all selection criteria, for the sum of the \FH and \SL channels. Minor background contributions are grouped into a single category labeled ``Other''. Also shown are several signal predictions in dashed colored lines for visualization only. The lower panels show the ``Pull'' defined as $\text{(Data - Prediction)/Total uncertainty}$.
    \label{fig:result:SR_SB_plots}
    }

\end{figure}
The analysis follows a blinded strategy, with all aspects of the event selection, background estimation, and statistical fitting procedure defined prior to unblinding data in the SR. The post-fit distributions in $M_{\PX}$ are shown in Fig. \ref{fig:result:SR_SB_plots} for the SR and SB. In the lower mass region, where the expected background yield is higher, we use a binning comparable to the expected signal width. For resonance masses above \(2~\TeV\), the background event counts become sparse. We therefore merge adjacent bins to avoid empty bins. The content of the final bin also includes events in the overflow region. The data are found to be in agreement with the SM background predictions within the uncertainties. We place upper limits on the production cross section for a heavy resonance \PX decaying to a pair of Higgs bosons.  

\begin{figure}[tbp]
\centering
   \includegraphics[width=0.48\textwidth]{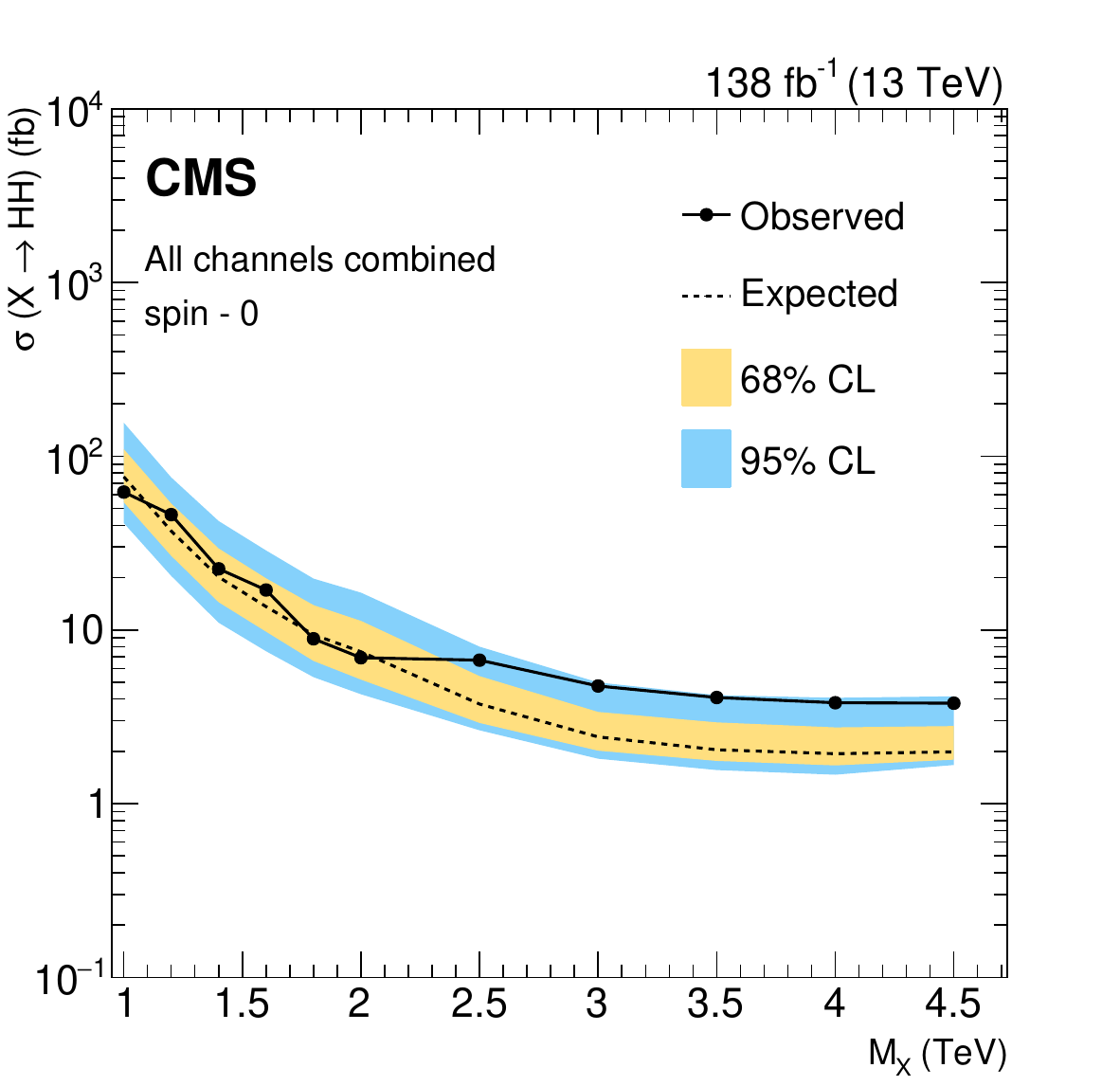}
   \includegraphics[width=0.48\textwidth]{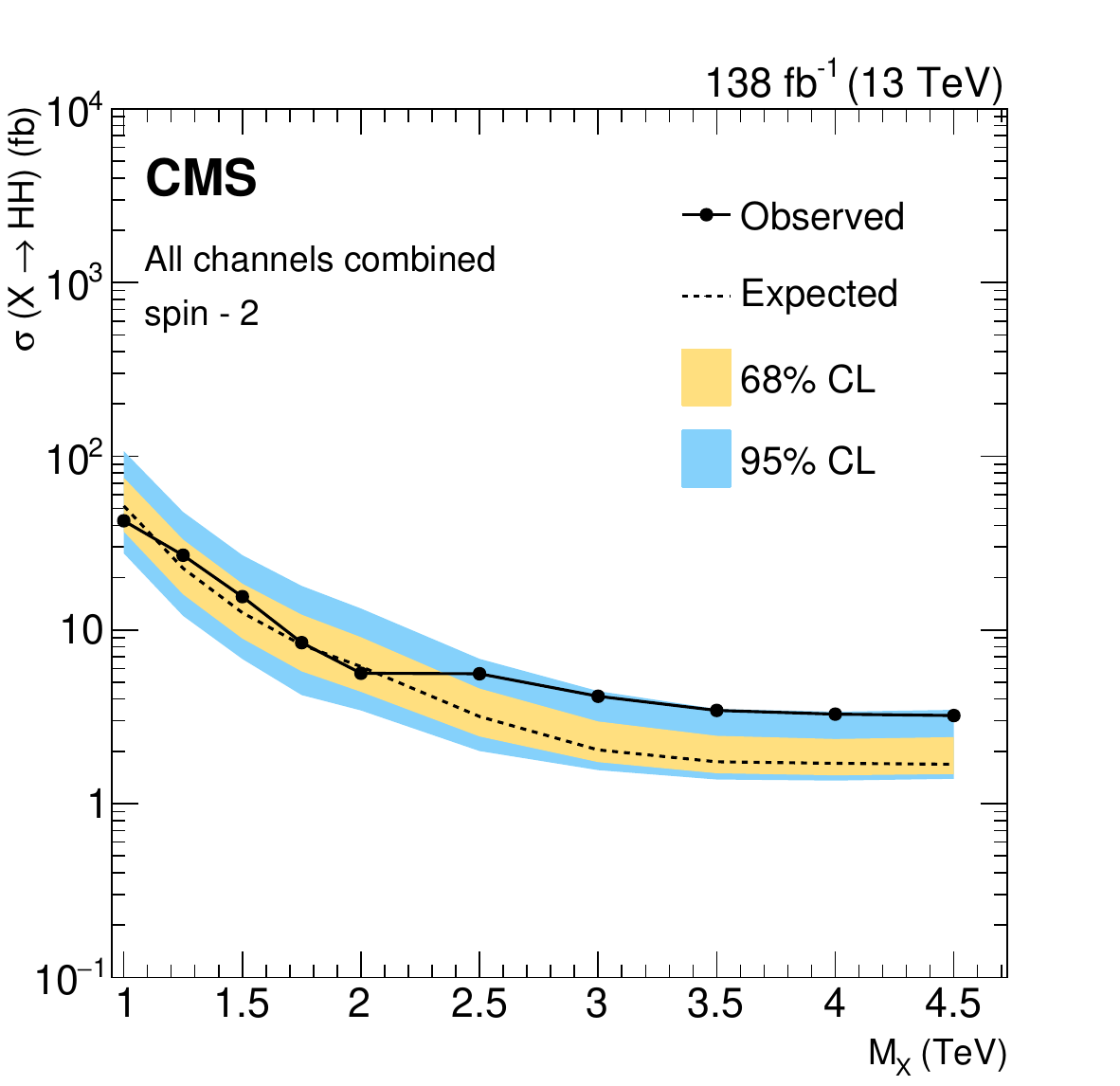}
   \caption{Expected and observed upper limits at 95\% CL on the production cross section of resonant \HH production for a spin-0 (\cmsLeft) and spin-2 (\cmsRight) narrow resonance. This calculation assumes SM branching fractions of the \PH boson as discussed in Section~\ref{sec:introduction}. The observed limits are higher than the expected limits for mass points above 2.5 TeV, primarily because of the excess of events visible in the highest bin of Fig.~\ref{fig:result:SR_SB_plots}~(\cmsLeft).
   \label{fig: unblinded_limits_run2}
   }
\end{figure}

The 95\% \CL upper limits on the $\PX \to \HH$ production for spin-0 and spin-2 resonances in the mass range from 1 to 4.5\TeV, computed using the modified frequentist \CL$_\text{s}$ method with a profile likelihood ratio test statistic, are shown in Fig.~\ref{fig: unblinded_limits_run2}. The distributions of the test statistic under the signal-plus-background and background-only hypotheses are obtained from ensembles of pseudo-experiments. The results are scaled by the SM branching fractions of the \PH bosons and the subsequent decays included in this analysis. The observed (expected) upper limits reach down to 62.2\unit{fb} (76.2\unit{fb}) for spin-0 and 42.5\unit{fb} (51.8\unit{fb}) for spin-2 at 1\TeV. 
At 4.5\TeV, the observed (expected) upper limits are 3.8\unit{fb} (2\unit{fb}) for spin-0 and 3.2\unit{fb} (1.7\unit{fb}) for spin-2. The results for spin-2 are broadly similar to those for spin-0. However, the \ptmiss spectrum for spin-2 signals is shifted toward higher values, which slightly increases the signal efficiency. This effect is visible in the superimposed signal templates for spin-0 and spin-2 in Fig.~\ref{fig:result:SR_SB_plots} and is reflected in the slightly stronger expected limits for the spin-2 scenario. The observed limits are higher than the expected limits for mass points above 2.5 TeV, primarily because of the excess of events visible in the highest bin of Fig.~\ref{fig:result:SR_SB_plots}~(\cmsLeft).

This analysis improves upon the spin-0 resolved ATLAS search~\cite{Run2atlasheavybbtautau} for resonance masses above 1.4\TeV. Furthermore, it is significantly better than the spin-0 search from ATLAS in the Lorentz-boosted regime~\cite{aad2020reconstruction}, covering the 1--3\TeV mass range, across the entire mass range considered in that analysis. The results presented in this paper represent the most sensitive upper limits to date in the resonance mass range of 1.4 to 4.5\TeV in the $\bbbar\tautau$ final state.

\section {Summary}
\label{sec:summary}
A search has been presented for heavy resonant Higgs boson pair (\HH) production in the $\bbbar\tautau$ final state, exploring resonance masses between 
1 and 4.5\TeV. The analysis is based on \pp collision data collected with the CMS detector during 2016--2018, corresponding to an integrated luminosity of 138\fbinv at a center-of-mass energy of 13\TeV. In this mass regime, the \PH bosons produced are boosted sufficiently to result in collimated decay products. 
The reconstruction and identification of such boosted objects are enhanced using advanced machine learning techniques, including a graph convolutional neural network for merged \bbbar jets and a convolutional neural network for boosted \tautau identification.

No significant deviation from the standard model background expectation is observed, and 95\% confidence level upper limits are set on the production cross section of a heavy resonance decaying to \HH, evaluated independently for both spin-0 and spin-2 hypotheses. This analysis sets the most sensitive upper bounds to date on the production of $\PX \to \HH \to \bbbar\tautau$ in the mass range of 1.4 to 4.5\TeV.

\begin{acknowledgments}
We congratulate our colleagues in the CERN accelerator departments for the excellent performance of the LHC and thank the technical and administrative staffs at CERN and at other CMS institutes for their contributions to the success of the CMS effort. In addition, we gratefully acknowledge the computing centers and personnel of the Worldwide LHC Computing Grid and other centers for delivering so effectively the computing infrastructure essential to our analyses. Finally, we acknowledge the enduring support for the construction and operation of the LHC, the CMS detector, and the supporting computing infrastructure provided by the following funding agencies: SC (Armenia), BMBWF and FWF (Austria); FNRS and FWO (Belgium); CNPq, CAPES, FAPERJ, FAPERGS, and FAPESP (Brazil); MES and BNSF (Bulgaria); CERN; CAS, MoST, and NSFC (China); MINCIENCIAS (Colombia); MSES and CSF (Croatia); RIF (Cyprus); SENESCYT (Ecuador); ERC PRG, TARISTU24-TK10 and MoER TK202 (Estonia); Academy of Finland, MEC, and HIP (Finland); CEA and CNRS/IN2P3 (France); SRNSF (Georgia); BMFTR, DFG, and HGF (Germany); GSRI (Greece); NKFIH (Hungary); DAE and DST (India); IPM (Iran); SFI (Ireland); INFN (Italy); MSIT and NRF (Republic of Korea); MES (Latvia); LMTLT (Lithuania); MOE and UM (Malaysia); BUAP, CINVESTAV, CONACYT, LNS, SEP, and UASLP-FAI (Mexico); MOS (Montenegro); MBIE (New Zealand); PAEC (Pakistan); MES, NSC, and NAWA (Poland); FCT (Portugal);  MESTD (Serbia); MICIU/AEI and PCTI (Spain); MOSTR (Sri Lanka); Swiss Funding Agencies (Switzerland); MST (Taipei); MHESI (Thailand); TUBITAK and TENMAK (T\"{u}rkiye); NASU (Ukraine); STFC (United Kingdom); DOE and NSF (USA).

\hyphenation{Rachada-pisek} Individuals have received support from the Marie-Curie program and the European Research Council and Horizon 2020 Grant, contract Nos.\ 675440, 724704, 752730, 758316, 765710, 824093, 101115353, 101002207, 101001205, and COST Action CA16108 (European Union); the Leventis Foundation; the Alfred P.\ Sloan Foundation; the Alexander von Humboldt Foundation; the Science Committee, project no. 22rl-037 (Armenia); the Fonds pour la Formation \`a la Recherche dans l'Industrie et dans l'Agriculture (FRIA) and Fonds voor Wetenschappelijk Onderzoek contract No. 1228724N (Belgium); the Beijing Municipal Science \& Technology Commission, No. Z191100007219010, the Fundamental Research Funds for the Central Universities, the Ministry of Science and Technology of China under Grant No. 2023YFA1605804, the Natural Science Foundation of China under Grant No. 12535004, and USTC Research Funds of the Double First-Class Initiative No.\ YD2030002017 (China); the Ministry of Education, Youth and Sports (MEYS) of the Czech Republic; the Shota Rustaveli National Science Foundation, grant FR-22-985 (Georgia); the Deutsche Forschungsgemeinschaft (DFG), among others, under Germany's Excellence Strategy -- EXC 2121 ``Quantum Universe" -- 390833306, and under project number 400140256 - GRK2497; the Hellenic Foundation for Research and Innovation (HFRI), Project Number 2288 (Greece); the Hungarian Academy of Sciences, the New National Excellence Program - \'UNKP, the NKFIH research grants K 131991, K 133046, K 138136, K 143460, K 143477, K 146913, K 146914, K 147048, 2020-2.2.1-ED-2021-00181, TKP2021-NKTA-64, and 2025-1.1.5-NEMZ\_KI-2025-00004 (Hungary); the Council of Science and Industrial Research, India; ICSC -- National Research Center for High Performance Computing, Big Data and Quantum Computing, FAIR -- Future Artificial Intelligence Research, and CUP I53D23001070006 (Mission 4 Component 1), funded by the NextGenerationEU program (Italy); the Latvian Council of Science; the Ministry of Education and Science, project no. 2022/WK/14, and the National Science Center, contracts Opus 2021/41/B/ST2/01369, 2021/43/B/ST2/01552, 2023/49/B/ST2/03273, and the NAWA contract BPN/PPO/2021/1/00011 (Poland); the Funda\c{c}\~ao para a Ci\^encia e a Tecnologia, grant CEECIND/01334/2018 (Portugal); the National Priorities Research Program by Qatar National Research Fund; MICIU/AEI/10.13039/501100011033, ERDF/EU, ``European Union NextGenerationEU/PRTR", and Programa Severo Ochoa del Principado de Asturias (Spain); the Chulalongkorn Academic into Its 2nd Century Project Advancement Project, the National Science, Research and Innovation Fund program IND\_FF\_68\_369\_2300\_097, and the Program Management Unit for Human Resources \& Institutional Development, Research and Innovation, grant B39G680009 (Thailand); the Kavli Foundation; the Nvidia Corporation; the SuperMicro Corporation; the Welch Foundation, contract C-1845; and the Weston Havens Foundation (USA).
\end{acknowledgments}

\bibliography{auto_generated}

\providecommand{\href}[2]{#2}\begingroup\raggedright\begin{thebibliography}{10}%
\makeatletter
\providecommand{\hrefCMSnoop }[0]{\@secondoftwo}%
\makeatother
\providecommand{\doi}{\texttt{doi:}\begingroup \urlstyle{tt}\Url}

\bibitem{CMS:2012qbp}
\hrefCMSnoop {}{{CMS Collaboration}, ``Observation of a new boson at a mass of
  {125\GeV} with the {CMS} experiment at the {LHC}'',} \textit{ Phys. Lett. B}
  \textbf{ 716} (2012) 30,
  \href{http://dx.doi.org/10.1016/j.physletb.2012.08.021}{\doi{10.1016/j.physletb.2012.08.021}},
  \href{http://www.arXiv.org/abs/1207.7235}{\texttt{arXiv:1207.7235}}.

\bibitem{CMS:2013btf}
\hrefCMSnoop {}{{CMS Collaboration}, ``Observation of a new boson with mass
  near 125 {GeV} in {$\Pp\Pp$} collisions at $\sqrt{s}$ = 7 and 8 {TeV}'',}
  \textit{ JHEP} \textbf{ 06} (2013) 081,
  \href{http://dx.doi.org/10.1007/JHEP06(2013)081}{\doi{10.1007/JHEP06(2013)081}},
  \href{http://www.arXiv.org/abs/1303.4571}{\texttt{arXiv:1303.4571}}.

\bibitem{ATLAS:2012yve}
\hrefCMSnoop {}{{ATLAS Collaboration}, ``Observation of a new particle in the
  search for the standard model {Higgs} boson with the {ATLAS} detector at the
  {LHC}'',} \textit{ Phys. Lett. B} \textbf{ 716} (2012) 1,
  \href{http://dx.doi.org/10.1016/j.physletb.2012.08.020}{\doi{10.1016/j.physletb.2012.08.020}},
  \href{http://www.arXiv.org/abs/1207.7214}{\texttt{arXiv:1207.7214}}.

\bibitem{ATLAS:2013xga}
\hrefCMSnoop {}{{ATLAS Collaboration}, ``{Evidence for the spin-0 nature of the
  Higgs boson using ATLAS data}'',} \textit{ Phys. Lett. B} \textbf{ 726}
  (2013) 120,
  \href{http://dx.doi.org/10.1016/j.physletb.2013.08.026}{\doi{10.1016/j.physletb.2013.08.026}},
  \href{http://www.arXiv.org/abs/1307.1432}{\texttt{arXiv:1307.1432}}.

\bibitem{PESKIN2025116971}
\hrefCMSnoop {}{M.~E. Peskin, ``{What is the Hierarchy Problem?}'',} \textit{
  Nucl. Phys. B} \textbf{ 1018} (2025) 116971,
  \href{http://dx.doi.org/10.1016/j.nuclphysb.2025.116971}{\doi{10.1016/j.nuclphysb.2025.116971}},
  \href{http://www.arXiv.org/abs/2505.00694}{\texttt{arXiv:2505.00694}}.

\bibitem{randall1999large}
\hrefCMSnoop {}{L.~Randall and R.~Sundrum, ``A large mass hierarchy from a
  small extra dimension'',} \textit{ Phys. Rev. Lett.} \textbf{ 83} (1999)
  3370,
  \href{http://dx.doi.org/10.1103/PhysRevLett.83.3370}{\doi{10.1103/PhysRevLett.83.3370}},
  \href{http://www.arXiv.org/abs/hep-ph/9905221}{\texttt{arXiv:hep-ph/9905221}}.

\bibitem{randall1999alternative}
\hrefCMSnoop {}{L.~Randall and R.~Sundrum, ``An alternative to
  compactification'',} \textit{ Phys. Rev. Lett.} \textbf{ 83} (1999) 4690,
  \href{http://dx.doi.org/10.1103/PhysRevLett.83.4690}{\doi{10.1103/PhysRevLett.83.4690}},
  \href{http://www.arXiv.org/abs/hep-th/9906064}{\texttt{arXiv:hep-th/9906064}}.

\bibitem{goldberger1999modulus}
\hrefCMSnoop {}{W.~D. Goldberger and M.~B. Wise, ``Modulus stabilization with
  bulk fields'',} \textit{ Phys. Rev. Lett.} \textbf{ 83} (1999) 4922,
  \href{http://dx.doi.org/10.1103/PhysRevLett.83.4922}{\doi{10.1103/PhysRevLett.83.4922}},
  \href{http://www.arXiv.org/abs/hep-ph/9907447}{\texttt{arXiv:hep-ph/9907447}}.

\bibitem{dewolfe2000modeling}
\hrefCMSnoop {}{O.~DeWolfe, D.~Freedman, S.~S. Gubser, and A.~Karch, ``Modeling
  the fifth dimension with scalars and gravity'',} \textit{ Phys. Rev. D}
  \textbf{ 62} (2000) 046008,
  \href{http://dx.doi.org/10.1103/PhysRevD.62.046008}{\doi{10.1103/PhysRevD.62.046008}},
  \href{http://www.arXiv.org/abs/hep-th/9909134}{\texttt{arXiv:hep-th/9909134}}.

\bibitem{csaki2000cosmology}
\hrefCMSnoop {}{C.~Csaki, M.~Graesser, L.~Randall, and J.~Terning, ``Cosmology
  of brane models with radion stabilization'',} \textit{ Phys. Rev. D} \textbf{
  62} (2000) 045015,
  \href{http://dx.doi.org/10.1103/PhysRevD.62.045015}{\doi{10.1103/PhysRevD.62.045015}},
  \href{http://www.arXiv.org/abs/hep-ph/9911406}{\texttt{arXiv:hep-ph/9911406}}.

\bibitem{davoudiasl2000phenomenology}
\hrefCMSnoop {}{H.~Davoudiasl, J.~Hewett, and T.~Rizzo, ``Phenomenology of the
  {R}andall--{S}undrum gauge hierarchy model'',} \textit{ Phys. Rev. Lett.}
  \textbf{ 84} (2000) 2080,
  \href{http://dx.doi.org/10.1103/PhysRevLett.84.2080}{\doi{10.1103/PhysRevLett.84.2080}},
  \href{http://www.arXiv.org/abs/hep-ph/9909255}{\texttt{arXiv:hep-ph/9909255}}.

\bibitem{agashe2007warped}
\hrefCMSnoop {}{K.~Agashe, H.~Davoudiasl, G.~Perez, and A.~Soni, ``Warped
  gravitons at the {CERN LHC} and beyond'',} \textit{ Phys. Rev. D} \textbf{
  76} (2007) 036006,
  \href{http://dx.doi.org/10.1103/PhysRevD.76.036006}{\doi{10.1103/PhysRevD.76.036006}},
  \href{http://www.arXiv.org/abs/hep-ph/0701186}{\texttt{arXiv:hep-ph/0701186}}.

\bibitem{2016cmsheavybbtautau}
\hrefCMSnoop {}{{CMS Collaboration}, ``Search for heavy resonances decaying
  into two {Higgs} bosons or into a {Higgs} boson and a {\PW} or {\PZ} boson in
  proton-proton collisions at {13\TeV}'',} \textit{ JHEP} \textbf{ 01} (2019)
  051,
  \href{http://dx.doi.org/10.1007/JHEP01(2019)051}{\doi{10.1007/JHEP01(2019)051}},
  \href{http://www.arXiv.org/abs/1808.01365}{\texttt{arXiv:1808.01365}}.

\bibitem{aad2020reconstruction}
\hrefCMSnoop {}{{ATLAS Collaboration}, ``Reconstruction and identification of
  boosted di-{$\PGt$} systems in a search for {H}iggs boson pairs using
  {13\TeV} proton-proton collision data in {ATLAS}'',} \textit{ JHEP} \textbf{
  11} (2020) 163,
  \href{http://dx.doi.org/10.1007/JHEP11(2020)163}{\doi{10.1007/JHEP11(2020)163}},
  \href{http://www.arXiv.org/abs/2007.14811}{\texttt{arXiv:2007.14811}}.

\bibitem{Run2atlasheavybbtautau}
\hrefCMSnoop {}{{ATLAS Collaboration}, ``Search for resonant and non-resonant
  {Higgs} boson pair production in the $\bbbar{\PGtp\PGtm}$ decay channel using
  {13\TeV} ${\Pp\Pp}$ collision data from the {ATLAS} detector'',} \textit{
  JHEP} \textbf{ 07} (2023) 040,
  \href{http://dx.doi.org/10.1007/JHEP07(2023)040}{\doi{10.1007/JHEP07(2023)040}},
  \href{http://www.arXiv.org/abs/2209.10910}{\texttt{arXiv:2209.10910}}.

\bibitem{deFlorian:2016spz}
\hrefCMSnoop {}{{LHC Higgs Cross Section Working Group}, ``Handbook of {LHC}
  {H}iggs cross sections: 4. {D}eciphering the nature of the {H}iggs sector'',}
  CERN Report CERN-2017-002-M, 2016.
\newblock
  \href{http://dx.doi.org/10.23731/CYRM-2017-002}{\doi{10.23731/CYRM-2017-002}},
  \href{http://www.arXiv.org/abs/1610.07922}{\texttt{arXiv:1610.07922}}.

\bibitem{PDG2024}
\hrefCMSnoop {}{{Particle Data Group}, S.~Navas { et~al.}, ``Review of particle
  physics'',} \textit{ Phys. Rev. D} \textbf{ 110} (2024) 030001,
  \href{http://dx.doi.org/10.1103/PhysRevD.110.030001}{\doi{10.1103/PhysRevD.110.030001}}.

\bibitem{qu2020jet}
\hrefCMSnoop {}{H.~Qu and L.~Gouskos, ``Jet tagging via particle clouds'',}
  \textit{ Phys. Rev. D} \textbf{ 101} (2020) 056019,
  \href{http://dx.doi.org/10.1103/PhysRevD.101.056019}{\doi{10.1103/PhysRevD.101.056019}},
  \href{http://www.arXiv.org/abs/1902.08570}{\texttt{arXiv:1902.08570}}.

\bibitem{CMS:2015fjg}
\hrefCMSnoop {}{{CMS Collaboration}, ``Search for narrow high-mass resonances
  in proton-proton collisions at $\sqrt{s}={8\TeV}$ decaying to {\PZ} and
  {Higgs} bosons'',} \textit{ Phys. Lett. B} \textbf{ 748} (2015) 255,
  \href{http://dx.doi.org/10.1016/j.physletb.2015.07.011}{\doi{10.1016/j.physletb.2015.07.011}},
  \href{http://www.arXiv.org/abs/1502.04994}{\texttt{arXiv:1502.04994}}.

\bibitem{CMS:2022prd}
\hrefCMSnoop {}{{CMS Collaboration}, ``Identification of hadronic tau lepton
  decays using a deep neural network'',} \textit{ JINST} \textbf{ 17} (2022)
  P07023,
  \href{http://dx.doi.org/10.1088/1748-0221/17/07/P07023}{\doi{10.1088/1748-0221/17/07/P07023}},
  \href{http://www.arXiv.org/abs/2201.08458}{\texttt{arXiv:2201.08458}}.

\bibitem{hepdata}
\hrefCMSnoop {}{}{HEPD}ata record for this analysis, 2026.
\newblock
  \href{http://dx.doi.org/10.17182/hepdata.166218}{\doi{10.17182/hepdata.166218}}.

\bibitem{Chatrchyan:2008zzk}
\hrefCMSnoop {}{{CMS Collaboration}, ``The {CMS} experiment at the {CERN}
  {LHC}'',} \textit{ JINST} \textbf{ 3} (2008) S08004,
  \href{http://dx.doi.org/10.1088/1748-0221/3/08/S08004}{\doi{10.1088/1748-0221/3/08/S08004}}.

\bibitem{CMS:2023gfb}
\hrefCMSnoop {}{{CMS Collaboration}, ``Development of the {CMS} detector for
  the {CERN LHC Run 3}'',} \textit{ JINST} \textbf{ 19} (2024) P05064,
  \href{http://dx.doi.org/10.1088/1748-0221/19/05/P05064}{\doi{10.1088/1748-0221/19/05/P05064}},
  \href{http://www.arXiv.org/abs/2309.05466}{\texttt{arXiv:2309.05466}}.

\bibitem{CMS:2016ngn}
\hrefCMSnoop {}{{CMS Collaboration}, ``The {CMS} trigger system'',} \textit{
  JINST} \textbf{ 12} (2017) P01020,
  \href{http://dx.doi.org/10.1088/1748-0221/12/01/P01020}{\doi{10.1088/1748-0221/12/01/P01020}},
\href{http://www.arXiv.org/abs/1609.02366}{\texttt{arXiv:1609.02366}}.

\bibitem{CMS:2020cmk}
\hrefCMSnoop {}{{CMS Collaboration}, ``Performance of the {CMS Level-1} trigger
  in proton-proton collisions at $\sqrt{s} = 13$ {TeV}'',} \textit{ JINST}
  \textbf{ 15} (2020) P10017,
  \href{http://dx.doi.org/10.1088/1748-0221/15/10/P10017}{\doi{10.1088/1748-0221/15/10/P10017}},
  \href{http://www.arXiv.org/abs/2006.10165}{\texttt{arXiv:2006.10165}}.

\bibitem{CMS:2024aqx}
\hrefCMSnoop {}{{CMS Collaboration}, ``Performance of the {CMS} high-level
  trigger during {LHC} {Run} 2'',} \textit{ JINST} \textbf{ 19} (2024) P11021,
  \href{http://dx.doi.org/10.1088/1748-0221/19/11/P11021}{\doi{10.1088/1748-0221/19/11/P11021}},
  \href{http://www.arXiv.org/abs/2410.17038}{\texttt{arXiv:2410.17038}}.

\bibitem{CMS-TDR-15-02}
\hrefCMSnoop {}{{CMS Collaboration}, ``Technical proposal for the {Phase-II}
  upgrade of the {Compact Muon Solenoid}'',} CMS Technical Proposal
  CERN-LHCC-2015-010, CMS-TDR-15-02, 2015.
\newblock
  \href{http://dx.doi.org/10.17181/CERN.VU8I.D59J}{\doi{10.17181/CERN.VU8I.D59J}}.

\bibitem{Frixione:2007vw}
\hrefCMSnoop {}{S.~Frixione, P.~Nason, and C.~Oleari, ``Matching {NLO} {QCD}
  computations with parton shower simulations: the {\POWHEG} method'',}
  \textit{ JHEP} \textbf{ 11} (2007) 070,
  \href{http://dx.doi.org/10.1088/1126-6708/2007/11/070}{\doi{10.1088/1126-6708/2007/11/070}},
  \href{http://www.arXiv.org/abs/0709.2092}{\texttt{arXiv:0709.2092}}.

\bibitem{Alioli:2010xd}
\hrefCMSnoop {}{S.~Alioli, P.~Nason, C.~Oleari, and E.~Re, ``A general
  framework for implementing {NLO} calculations in shower {Monte Carlo}
  programs: the {\POWHEG} \textsc{box}'',} \textit{ JHEP} \textbf{ 06} (2010)
  043,
  \href{http://dx.doi.org/10.1007/JHEP06(2010)043}{\doi{10.1007/JHEP06(2010)043}},
  \href{http://www.arXiv.org/abs/1002.2581}{\texttt{arXiv:1002.2581}}.

\bibitem{Czakon:2017wor}
M.~Czakon\hrefCMSnoop {}{ { et~al.}, ``Top-pair production at the {LHC} through
  {NNLO} {QCD} and {NLO} {EW}'',} \textit{ JHEP} \textbf{ 10} (2017) 186,
  \href{http://dx.doi.org/10.1007/JHEP10(2017)186}{\doi{10.1007/JHEP10(2017)186}},
  \href{http://www.arXiv.org/abs/1705.04105}{\texttt{arXiv:1705.04105}}.

\bibitem{Alwall:2014hca}
J.~Alwall\hrefCMSnoop {}{ { et~al.}, ``The automated computation of tree-level
  and next-to-leading order differential cross sections, and their matching to
  parton shower simulations'',} \textit{ JHEP} \textbf{ 07} (2014) 079,
  \href{http://dx.doi.org/10.1007/JHEP07(2014)079}{\doi{10.1007/JHEP07(2014)079}},
  \href{http://www.arXiv.org/abs/1405.0301}{\texttt{arXiv:1405.0301}}.

\bibitem{Mangano:2006rw}
\hrefCMSnoop {}{M.~L. Mangano, M.~Moretti, F.~Piccinini, and M.~Treccani,
  ``Matching matrix elements and shower evolution for top-pair production in
  hadronic collisions'',} \textit{ JHEP} \textbf{ 01} (2007) 013,
  \href{http://dx.doi.org/10.1088/1126-6708/2007/01/013}{\doi{10.1088/1126-6708/2007/01/013}},
  \href{http://www.arXiv.org/abs/hep-ph/0611129}{\texttt{arXiv:hep-ph/0611129}}.

\bibitem{Denner:2009gj}
\hrefCMSnoop {}{A.~Denner, S.~Dittmaier, T.~Kasprzik, and A.~Muck,
  ``Electroweak corrections to {W+jet} hadroproduction including leptonic
  {W-boson} decays'',} \textit{ JHEP} \textbf{ 08} (2009) 075,
  \href{http://dx.doi.org/10.1088/1126-6708/2009/08/075}{\doi{10.1088/1126-6708/2009/08/075}},
\href{http://www.arXiv.org/abs/0906.1656}{\texttt{arXiv:0906.1656}}.

\bibitem{Denner:2011vu}
\hrefCMSnoop {}{A.~Denner, S.~Dittmaier, T.~Kasprzik, and A.~Muck,
  ``Electroweak corrections to dilepton+jet production at hadron colliders'',}
  \textit{ JHEP} \textbf{ 06} (2011) 069,
  \href{http://dx.doi.org/10.1007/JHEP06(2011)069}{\doi{10.1007/JHEP06(2011)069}},
\href{http://www.arXiv.org/abs/1103.0914}{\texttt{arXiv:1103.0914}}.

\bibitem{Denner:2012ts}
\hrefCMSnoop {}{A.~Denner, S.~Dittmaier, T.~Kasprzik, and A.~Maeck,
  ``Electroweak corrections to monojet production at the {LHC}'',} \textit{
  Eur. Phys. J. C} \textbf{ 73} (2013) 2297,
  \href{http://dx.doi.org/10.1140/epjc/s10052-013-2297-x}{\doi{10.1140/epjc/s10052-013-2297-x}},
\href{http://www.arXiv.org/abs/1211.5078}{\texttt{arXiv:1211.5078}}.

\bibitem{Kuhn:2005gv}
\hrefCMSnoop {}{J.~H. Kuhn, A.~Kulesza, S.~Pozzorini, and M.~Schulze,
  ``Electroweak corrections to hadronic photon production at large transverse
  momenta'',} \textit{ JHEP} \textbf{ 03} (2006) 059,
  \href{http://dx.doi.org/10.1088/1126-6708/2006/03/059}{\doi{10.1088/1126-6708/2006/03/059}},
\href{http://www.arXiv.org/abs/hep-ph/0508253}{\texttt{arXiv:hep-ph/0508253}}.

\bibitem{Kallweit:2014xda}
S.~Kallweit\hrefCMSnoop {}{ { et~al.}, ``{NLO} electroweak automation and
  precise predictions for {W+multijet} production at the {LHC}'',} \textit{
  JHEP} \textbf{ 04} (2015) 012,
  \href{http://dx.doi.org/10.1007/JHEP04(2015)012}{\doi{10.1007/JHEP04(2015)012}},
\href{http://www.arXiv.org/abs/1412.5157}{\texttt{arXiv:1412.5157}}.

\bibitem{Kallweit:2015dum}
S.~Kallweit\hrefCMSnoop {}{ { et~al.}, ``{NLO} {QCD}+{EW} predictions for
  {\PV}+jets including off-shell vector-boson decays and multijet merging'',}
  \textit{ JHEP} \textbf{ 04} (2016) 021,
  \href{http://dx.doi.org/10.1007/JHEP04(2016)021}{\doi{10.1007/JHEP04(2016)021}},
  \href{http://www.arXiv.org/abs/1511.08692}{\texttt{arXiv:1511.08692}}.

\bibitem{Sjostrand:2007gs}
T.~Sj{\"o}strand\hrefCMSnoop {}{ { et~al.}, ``An introduction to {PYTHIA}
  8.2'',} \textit{ Comput. Phys. Commun.} \textbf{ 191} (2015) 159,
  \href{http://dx.doi.org/10.1016/j.cpc.2015.01.024}{\doi{10.1016/j.cpc.2015.01.024}},
  \href{http://www.arXiv.org/abs/1410.3012}{\texttt{arXiv:1410.3012}}.

\bibitem{Alwall:2007fs}
J.~Alwall\hrefCMSnoop {}{ { et~al.}, ``Comparative study of various algorithms
  for the merging of parton showers and matrix elements in hadronic
  collisions'',} \textit{ Eur. Phys. J. C} \textbf{ 53} (2008) 473,
  \href{http://dx.doi.org/10.1140/epjc/s10052-007-0490-5}{\doi{10.1140/epjc/s10052-007-0490-5}},
  \href{http://www.arXiv.org/abs/0706.2569}{\texttt{arXiv:0706.2569}}.

\bibitem{Frederix:2012ps}
\hrefCMSnoop {}{R.~Frederix and S.~Frixione, ``Merging meets matching in
  {\MCATNLO}'',} \textit{ JHEP} \textbf{ 12} (2012) 061,
  \href{http://dx.doi.org/10.1007/JHEP12(2012)061}{\doi{10.1007/JHEP12(2012)061}},
  \href{http://www.arXiv.org/abs/1209.6215}{\texttt{arXiv:1209.6215}}.

\bibitem{Ball_2017}
\hrefCMSnoop {}{{NNPDF} Collaboration, ``Parton distributions from
  high-precision collider data'',} \textit{ Eur. Phys. J. C} \textbf{ 77}
  (2017) 663,
  \href{http://dx.doi.org/10.1140/epjc/s10052-017-5199-5}{\doi{10.1140/epjc/s10052-017-5199-5}},
  \href{http://www.arXiv.org/abs/1706.00428}{\texttt{arXiv:1706.00428}}.

\bibitem{CMS:2019csb}
\hrefCMSnoop {}{{CMS Collaboration}, ``Extraction and validation of a new set
  of {CMS} {\PYTHIA8} tunes from underlying-event measurements'',} \textit{
  Eur. Phys. J. C} \textbf{ 80} (2020) 4,
  \href{http://dx.doi.org/10.1140/epjc/s10052-019-7499-4}{\doi{10.1140/epjc/s10052-019-7499-4}},
  \href{http://www.arXiv.org/abs/1903.12179}{\texttt{arXiv:1903.12179}}.

\bibitem{Agostinelli:2002hh}
\hrefCMSnoop {}{{GEANT4} Collaboration, ``{\GEANTfour}---a simulation
  toolkit'',} \textit{ Nucl. Instrum. Meth. A} \textbf{ 506} (2003) 250,
  \href{http://dx.doi.org/10.1016/S0168-9002(03)01368-8}{\doi{10.1016/S0168-9002(03)01368-8}}.

\bibitem{Sirunyan:2017ulk}
\hrefCMSnoop {}{{CMS Collaboration}, ``Particle-flow reconstruction and global
  event description with the {CMS} detector'',} \textit{ JINST} \textbf{ 12}
  (2017) P10003,
  \href{http://dx.doi.org/10.1088/1748-0221/12/10/P10003}{\doi{10.1088/1748-0221/12/10/P10003}},
  \href{http://www.arXiv.org/abs/1706.04965}{\texttt{arXiv:1706.04965}}.

\bibitem{Cacciari:2008gp}
\hrefCMSnoop {}{M.~Cacciari, G.~P. Salam, and G.~Soyez, ``The anti-\kt jet
  clustering algorithm'',} \textit{ JHEP} \textbf{ 04} (2008) 063,
  \href{http://dx.doi.org/10.1088/1126-6708/2008/04/063}{\doi{10.1088/1126-6708/2008/04/063}},
  \href{http://www.arXiv.org/abs/0802.1189}{\texttt{arXiv:0802.1189}}.

\bibitem{Cacciari:2011ma}
\hrefCMSnoop {}{M.~Cacciari, G.~P. Salam, and G.~Soyez, ``{\FASTJET} user
  manual'',} \textit{ Eur. Phys. J. C} \textbf{ 72} (2012) 1896,
  \href{http://dx.doi.org/10.1140/epjc/s10052-012-1896-2}{\doi{10.1140/epjc/s10052-012-1896-2}},
  \href{http://www.arXiv.org/abs/1111.6097}{\texttt{arXiv:1111.6097}}.

\bibitem{CMS:2019ctu}
\hrefCMSnoop {}{{CMS Collaboration}, ``{Performance of missing transverse
  momentum reconstruction in proton-proton collisions at $\sqrt{s} =$ 13 TeV
  using the CMS detector}'',} \textit{ JINST} \textbf{ 14} (2019) P07004,
  \href{http://dx.doi.org/10.1088/1748-0221/14/07/P07004}{\doi{10.1088/1748-0221/14/07/P07004}},
  \href{http://www.arXiv.org/abs/1903.06078}{\texttt{arXiv:1903.06078}}.

\bibitem{Bertolini2014}
\hrefCMSnoop {}{D.~Bertolini, P.~Harris, M.~Low, and N.~Tran, ``Pileup per
  particle identification'',} \textit{ JHEP} \textbf{ 10} (2014) 059,
  \href{http://dx.doi.org/10.1007/JHEP10(2014)059}{\doi{10.1007/JHEP10(2014)059}},
  \href{http://www.arXiv.org/abs/1407.6013}{\texttt{arXiv:1407.6013}}.

\bibitem{CMS:2025kje}
\hrefCMSnoop {}{{CMS Collaboration}, ``{Performance of heavy-flavour jet
  identification in Lorentz-boosted topologies in proton-proton collisions at
  $\sqrt{s}=$ 13 TeV}'',} \textit{ JINST} \textbf{ 20} (2025) P11006,
  \href{http://dx.doi.org/10.1088/1748-0221/20/11/P11006}{\doi{10.1088/1748-0221/20/11/P11006}},
  \href{http://www.arXiv.org/abs/2510.10228}{\texttt{arXiv:2510.10228}}.

\bibitem{bols2020jet}
E.~Bols\hrefCMSnoop {}{ { et~al.}, ``Jet flavour classification using
  {DeepJet}'',} \textit{ JINST} \textbf{ 15} (2020) P12012,
  \href{http://dx.doi.org/10.1088/1748-0221/15/12/P12012}{\doi{10.1088/1748-0221/15/12/P12012}},
  \href{http://www.arXiv.org/abs/2008.10519}{\texttt{arXiv:2008.10519}}.

\bibitem{CMS:2018jrd}
\hrefCMSnoop {}{{CMS Collaboration}, ``{Performance of reconstruction and
  identification of $\tau$ leptons decaying to hadrons and $\nu_\tau$ in pp
  collisions at $\sqrt{s}=$ 13 TeV}'',} \textit{ JINST} \textbf{ 13} (2018)
  P10005,
  \href{http://dx.doi.org/10.1088/1748-0221/13/10/P10005}{\doi{10.1088/1748-0221/13/10/P10005}},
  \href{http://www.arXiv.org/abs/1809.02816}{\texttt{arXiv:1809.02816}}.

\bibitem{Wobisch:1998wt}
\href
  {http://inspirehep.net/record/484872/files/arXiv:hep-ph_9907280.pdf}{M.~Wobisch
  and T.~Wengler, ``{Hadronization corrections to jet cross-sections in deep
  inelastic scattering}'',} in \textit{ {Proc. Workshop on Monte Carlo
  generators for HERA physics, Hamburg, Germany}}.
\newblock 1998.
\newblock
\href{http://www.arXiv.org/abs/hep-ph/9907280}{\texttt{arXiv:hep-ph/9907280}}.
\newblock

\bibitem{Chatrchyan:2012zz}
\hrefCMSnoop {}{{CMS Collaboration}, ``Reconstruction and identification of
  {\PGt} lepton decays to hadrons and {\PGnGt} at {CMS}'',} \textit{ JINST}
  \textbf{ 7} (2012) P01001,
  \href{http://dx.doi.org/10.1088/1748-0221/7/01/P01001}{\doi{10.1088/1748-0221/7/01/P01001}},
  \href{http://www.arXiv.org/abs/1109.6034}{\texttt{arXiv:1109.6034}}.

\bibitem{CMS-DP-2025-047}
\href {https://cds.cern.ch/record/2941434}{{CMS Collaboration}, ``{Performance
  of boosted tau lepton identification with DeepTau Framework (Boosted
  DeepTau)}'',} {CMS Detector Performance Summary} CMS-DP-2025-047, 2025.

\bibitem{Khachatryan:2015hwa}
\hrefCMSnoop {}{{CMS Collaboration}, ``Performance of electron reconstruction
  and selection with the {CMS} detector in proton-proton collisions at
  $\sqrt{s}={8\TeV}$'',} \textit{ JINST} \textbf{ 10} (2015) P06005,
  \href{http://dx.doi.org/10.1088/1748-0221/10/06/P06005}{\doi{10.1088/1748-0221/10/06/P06005}},
  \href{http://www.arXiv.org/abs/1502.02701}{\texttt{arXiv:1502.02701}}.

\bibitem{Chatrchyan:2012xi}
\hrefCMSnoop {}{{CMS Collaboration}, ``Performance of {CMS} muon reconstruction
  in ${\Pp\Pp}$ collision events at $\sqrt{s}={7\TeV}$'',} \textit{ JINST}
  \textbf{ 7} (2012) P10002,
  \href{http://dx.doi.org/10.1088/1748-0221/7/10/P10002}{\doi{10.1088/1748-0221/7/10/P10002}},
  \href{http://www.arXiv.org/abs/1206.4071}{\texttt{arXiv:1206.4071}}.

\bibitem{Larkoski:2014wba}
\hrefCMSnoop {}{A.~J. Larkoski, S.~Marzani, G.~Soyez, and J.~Thaler, ``Soft
  drop'',} \textit{ JHEP} \textbf{ 05} (2014) 146,
  \href{http://dx.doi.org/10.1007/JHEP05(2014)146}{\doi{10.1007/JHEP05(2014)146}},
  \href{http://www.arXiv.org/abs/1402.2657}{\texttt{arXiv:1402.2657}}.

\bibitem{Matyszkiewicz:2025gal}
\hrefCMSnoop {}{W.~Matyszkiewicz and A.~Kalinowski, ``{Tau-pair invariant mass
  estimation using maximum likelihood estimation and collinear
  approximation}'',} \textit{ Acta Phys. Polon. Supp.} \textbf{ 18} (2025) 5,
  \href{http://dx.doi.org/10.5506/APhysPolBSupp.18.5-A21}{\doi{10.5506/APhysPolBSupp.18.5-A21}}.

\bibitem{Ellis:1987xu}
\hrefCMSnoop {}{R.~K. Ellis, I.~Hinchliffe, M.~Soldate, and J.~J. van~der Bij,
  ``{{H}iggs decay to {\tautau}: {A} possible signature of intermediate mass
  {H}iggs bosons at high energy hadron colliders}'',} \textit{ Nucl. Phys. B}
  \textbf{ 297} (1988) 221,
  \href{http://dx.doi.org/10.1016/0550-3213(88)90019-3}{\doi{10.1016/0550-3213(88)90019-3}}.

\bibitem{hayrapetyan2024cms}
\hrefCMSnoop {}{{CMS Collaboration}, ``The {CMS} statistical analysis and
  combination tool: \textsc{combine}'',} \textit{ Comput. Softw. Big Sci.}
  \textbf{ 8} (2024) 19,
  \href{http://dx.doi.org/10.1007/s41781-024-00121-4}{\doi{10.1007/s41781-024-00121-4}},
  \href{http://www.arXiv.org/abs/2404.06614}{\texttt{arXiv:2404.06614}}.

\bibitem{Verkerke:2003ir}
\href
  {https://www.slac.stanford.edu/econf/C0303241/proc/papers/MOLT007.PDF}{W.~Verkerke
  and D.~Kirkby, ``The \textsc{RooFit} toolkit for data modeling'',} in
  \textit{ {Proc. 13th In. Conf. on Computing in High Energy and Nuclear
  Physics (CHEP 2003): La Jolla CA, United States, March 24--28}}.
\newblock 2003.
\newblock
  \href{http://www.arXiv.org/abs/physics/0306116}{\texttt{arXiv:physics/0306116}}.

\bibitem{Moneta:2010pm}
L.~Moneta\hrefCMSnoop {}{ { et~al.}, ``The \textsc{RooStats} project'',} in
  \textit{ {Proc. 13th International Workshop on Advanced Computing and
  Analysis Techniques in Physics Research (ACAT 2010): Jaipur, India, February
  22--27}}.
\newblock 2010.
\newblock \href{http://www.arXiv.org/abs/1009.1003}{\texttt{arXiv:1009.1003}}.
\newblock [PoS (ACAT2010) 057].
  \href{http://dx.doi.org/10.22323/1.093.0057}{\doi{10.22323/1.093.0057}}.

\bibitem{CMS-LUM-17-003}
\hrefCMSnoop {}{{CMS Collaboration}, ``Precision luminosity measurement in
  proton-proton collisions at $\sqrt{s}={13\TeV}$ in 2015 and 2016 at {CMS}'',}
  \textit{ Eur. Phys. J. C} \textbf{ 81} (2021) 800,
  \href{http://dx.doi.org/10.1140/epjc/s10052-021-09538-2}{\doi{10.1140/epjc/s10052-021-09538-2}},
  \href{http://www.arXiv.org/abs/2104.01927}{\texttt{arXiv:2104.01927}}.

\bibitem{CMS-PAS-LUM-20-001}
\href {https://cds.cern.ch/record/2940794}{{CMS Collaboration}, ``{Precision
  luminosity measurement in proton-proton collisions at $\sqrt{s}={13\TeV}$
  with the CMS detector}'',} {CMS Physics Analysis Summary} CMS-PAS-LUM-20-001,
  2025.

\bibitem{Ball:2014uwa}
\hrefCMSnoop {}{{NNPDF} Collaboration, ``Parton distributions for the {LHC} run
  {II}'',} \textit{ JHEP} \textbf{ 04} (2015) 040,
  \href{http://dx.doi.org/10.1007/JHEP04(2015)040}{\doi{10.1007/JHEP04(2015)040}},
  \href{http://www.arXiv.org/abs/1410.8849}{\texttt{arXiv:1410.8849}}.

\bibitem{Butterworth:2015oua}
\hrefCMSnoop {}{J.~Butterworth { et~al.}, ``{PDF4LHC recommendations for LHC
  Run II}'',} \textit{ J. Phys. G} \textbf{ 43} (2016) 023001,
  \href{http://dx.doi.org/10.1088/0954-3899/43/2/023001}{\doi{10.1088/0954-3899/43/2/023001}},
  \href{http://www.arXiv.org/abs/1510.03865}{\texttt{arXiv:1510.03865}}.

\bibitem{Sirunyan:2018nqx}
\hrefCMSnoop {}{{CMS Collaboration}, ``Measurement of the inelastic
  proton-proton cross section at $\sqrt{s}={13\TeV}$'',} \textit{ JHEP}
  \textbf{ 07} (2018) 161,
  \href{http://dx.doi.org/10.1007/JHEP07(2018)161}{\doi{10.1007/JHEP07(2018)161}},
  \href{http://www.arXiv.org/abs/1802.02613}{\texttt{arXiv:1802.02613}}.

\bibitem{Khachatryan:2016kdb}
\hrefCMSnoop {}{{CMS Collaboration}, ``Jet energy scale and resolution in the
  {CMS} experiment in ${\Pp\Pp}$ collisions at {8\TeV}'',} \textit{ JINST}
  \textbf{ 12} (2017) P02014,
  \href{http://dx.doi.org/10.1088/1748-0221/12/02/P02014}{\doi{10.1088/1748-0221/12/02/P02014}},
  \href{http://www.arXiv.org/abs/1607.03663}{\texttt{arXiv:1607.03663}}.

\bibitem{Sirunyan:2017ezt}
\hrefCMSnoop {}{{CMS Collaboration}, ``Identification of heavy-flavour jets
  with the {CMS} detector in ${\Pp\Pp}$ collisions at {13\TeV}'',} \textit{
  JINST} \textbf{ 13} (2018) P05011,
  \href{http://dx.doi.org/10.1088/1748-0221/13/05/P05011}{\doi{10.1088/1748-0221/13/05/P05011}},
  \href{http://www.arXiv.org/abs/1712.07158}{\texttt{arXiv:1712.07158}}.

\bibitem{BARLOW1993219}
\hrefCMSnoop {}{R.~Barlow and C.~Beeston, ``Fitting using finite {Monte Carlo}
  samples'',} \textit{ Comput. Phys. Commun.} \textbf{ 77} (1993) 219,
  \href{http://dx.doi.org/10.1016/0010-4655(93)90005-W}{\doi{10.1016/0010-4655(93)90005-W}}.

\end{thebibliography}\endgroup

\cleardoublepage \appendix\section{The CMS Collaboration \label{app:collab}}\begin{sloppypar}\hyphenpenalty=5000\widowpenalty=500\clubpenalty=5000\cmsinstitute{Yerevan Physics Institute, Yerevan, Armenia}
{\tolerance=6000
A.~Hayrapetyan, V.~Makarenko\cmsorcid{0000-0002-8406-8605}, A.~Tumasyan\cmsAuthorMark{1}\cmsorcid{0009-0000-0684-6742}
\par}
\cmsinstitute{Institut f\"{u}r Hochenergiephysik, Vienna, Austria}
{\tolerance=6000
W.~Adam\cmsorcid{0000-0001-9099-4341}, L.~Benato\cmsorcid{0000-0001-5135-7489}, T.~Bergauer\cmsorcid{0000-0002-5786-0293}, M.~Dragicevic\cmsorcid{0000-0003-1967-6783}, P.S.~Hussain\cmsorcid{0000-0002-4825-5278}, M.~Jeitler\cmsAuthorMark{2}\cmsorcid{0000-0002-5141-9560}, N.~Krammer\cmsorcid{0000-0002-0548-0985}, A.~Li\cmsorcid{0000-0002-4547-116X}, D.~Liko\cmsorcid{0000-0002-3380-473X}, M.~Matthewman, J.~Schieck\cmsAuthorMark{2}\cmsorcid{0000-0002-1058-8093}, R.~Sch\"{o}fbeck\cmsAuthorMark{2}\cmsorcid{0000-0002-2332-8784}, M.~Shooshtari\cmsorcid{0009-0004-8882-4887}, M.~Sonawane\cmsorcid{0000-0003-0510-7010}, W.~Waltenberger\cmsorcid{0000-0002-6215-7228}, C.-E.~Wulz\cmsAuthorMark{2}\cmsorcid{0000-0001-9226-5812}
% ...
\par}
\vskip\cmsinstskip
\dag:~Deceased\\
$^{1}$Also at Yerevan State University, Yerevan, Armenia\\
$^{2}$Also at TU Wien, Vienna, Austria\\
% ...
\end{sloppypar}
\end{document}